\documentclass[twocolumn,prd,superscriptaddress,nofootinbib,preprintnumbers]{revtex4-1}
\usepackage{graphicx}
\usepackage{amsmath,bm,amssymb,amsfonts,dsfont,xspace}
\usepackage{textgreek}
\usepackage{derivative}
\usepackage[usenames,dvipsnames]{xcolor}
\usepackage[normalem]{ulem}
\usepackage{url}
\usepackage{footmisc}
\usepackage{multirow}
\usepackage[colorlinks  = true,
            linkcolor   = NavyBlue,
            urlcolor    = NavyBlue,
            citecolor   = NavyBlue,
            anchorcolor = NavyBlue, unicode]{hyperref}

\newcommand{\lsim}{\mathrel{\mathop{\kern 0pt \rlap
  {\raise.2ex\hbox{$<$}}}
  \lower.9ex\hbox{\kern-.190em $\approx$}}}
\newcommand{\gsim}{\mathrel{\mathop{\kern 0pt \rlap
  {\raise.2ex\hbox{$>$}}}
  \lower.9ex\hbox{\kern-.190em $\approx$}}}

\newcommand{\vMol}{v}
\newcommand{\sigmav}{$\langle\sigma \vMol\rangle$ }
  
\renewcommand{\vec}[1]{\boldsymbol{#1}}

\newcommand{\maddm}{{\sc MadDM}\xspace}
\newcommand{\mg}{{\sc MG5\_aMC}\xspace}
\newcommand{\micromegas}{{\sc micrOMEGAs}\xspace}

\newcommand{\pythia}{{\sc Pythia}\xspace}
\newcommand{\drake}{{\sc DRAKE}\xspace}
\newcommand{\python}{{\sc Python}\xspace}
\newcommand{\github}{{\sc Github}\xspace}
\newcommand{\singlet}{{\sc 
SingletScalar\_DM}\xspace}
  
\usepackage[normalem]{ulem} %temporary

\def \aap  {A\&A}

\def \apjs  {ApJS}
\def \apjl  {ApJL}

\begin{document}

%\preprint{TTK-20-XX}

\title{Dark matter in the Higgs resonance region}

\author{Mattia Di Mauro}\email{dimauro.mattia@gmail.com}
\affiliation{Istituto Nazionale di Fisica Nucleare, Sezione di Torino, Via P. Giuria 1, 10125 Torino, Italy}

\author{Chiara Arina}\email{chiara.arina@uclouvain.be}
\affiliation{Centre for Cosmology, Particle Physics and Phenomenology (CP3), Universit\'e catholique de Louvain, Chemin du Cyclotron 2, 1348 Louvain-la-Neuve, Belgium}

\author{Nicolao Fornengo}\email{nicolao.fornengo@unito.it}
\affiliation{Department of Physics, University of Torino, Via P. Giuria 1, 10125 Torino, Italy}
\affiliation{Istituto Nazionale di Fisica Nucleare, Sezione di Torino, Via P. Giuria 1, 10125 Torino, Italy}

\author{Jan Heisig}\email{heisig@virginia.edu}
\affiliation{Department of Physics, University of Virginia, Charlottesville, Virginia 22904-4714, USA}

\author{Daniele Massaro}\email{daniele.massaro@uclouvain.be}
\affiliation{Dipartimento di Fisica e Astronomia, Alma Mater Studiorum - Universit\`a di Bologna, via Irnerio 46, 40126 Bologna, Italy}
\affiliation{Istituto Nazionale di Fisica Nucleare, Sezione di Bologna, viale Berti Pichat 6/2, 40127 Bologna, Italy}
\affiliation{Centre for Cosmology, Particle Physics and Phenomenology (CP3), Universit\'e catholique de Louvain, Chemin du Cyclotron 2, 1348 Louvain-la-Neuve, Belgium}

%\date{\today} 
\begin{abstract}
The singlet scalar Higgs portal model 
provides one of the simplest explanations of dark matter in our Universe. 
Its Higgs resonant region, $m_\text{DM}\approx m_h/2$, has gained particular attention, being able to reconcile the tension between the relic density measurement and direct detection constraints. Interestingly, this region is also preferred as an explanation of the Fermi-LAT  $\gamma$-ray Galactic center excess. We perform a 
detailed study of this 
model using $\gamma$-ray data from the Galactic center and from
dwarf spheroidal galaxies and combine them with cosmic-ray antiproton data from the AMS-02 experiment that shows a compatible excess. 
In the calculation of the relic density, we take into account effects of early kinetic decoupling relevant for resonant annihilation. 
The model provides excellent fits to the astrophysical data either in the case the dark matter candidate constitutes all or a subdominant fraction of the observed relic density.
We show projections for future direct detection and collider experiments to probe these scenarios.
\end{abstract}

\maketitle
%\flushbottom

%%%%%%%%%%%%%%%%%%%%%%%%%%%%%%%%%%%%%%%%%%%%%%%%%%%%%%%%%%%%
\section{Introduction}
\label{sec:intro}

The origin of dark matter (DM) is one of the main unsolved puzzles in fundamental physics today, implying physics Beyond the Standard Model (BSM).
The Singlet Scalar Higgs portal (SHP) model is among the
simplest UV complete BSM theories that provide a valid DM candidate~\cite{SILVEIRA1985136,McDonald:1993ex,Burgess:2000yq,Davoudiasl:2004be,Ham:2004cf,OConnell:2006rsp,Profumo:2007wc,Barger:2007im,Lerner:2009xg}. While a large portion of the model parameter space has already been excluded by the interplay of direct detection and relic density constraints, the Higgs resonant region, $m_{\rm{DM}}\approx m_h/2$, is particularly challenging to probe. It constitutes one of the two remaining viable regions~\cite{GAMBIT:2017gge,Athron:2018ipf} within the model. Independently, a DM mass of around 60\,GeV is also preferred as an explanation of two potential indirect detection signals of DM~\cite{Cuoco:2017rxb} as we detail below. In fact, putting together all relevant observational data allows for a scenario where the DM candidate constitutes all or just a subdominant fraction of the observed DM density and reveals an intricate relation between the model parameters and the DM fraction. Due to the sharp resonance, the results are highly sensitive to the DM mass in the region $m_\text{DM} = m_h/2 \pm \mathcal{O}(\Gamma_{h})$, where $\Gamma_{h}$ denotes the total Higgs width. Therefore, the Higgs resonant region deserves a closer look, which we will provide in this paper.

Several groups have reported the detection of an excess of $\gamma$ rays, which is labeled as the Galactic center excess (GCE), in the data of the {\it Fermi} Large Area Telescope ({\it Fermi}-LAT) in the direction of the center of the Milky Way (see, e.g., \cite{Goodenough:2009gk,Hooper:2010mq,Boyarsky:2010dr,Hooper:2011ti,Abazajian:2012pn,Gordon:2013vta,Abazajian:2014fta,Daylan:2014rsa,Calore:2014nla,Calore:2014xka,TheFermi-LAT:2015kwa,Cuoco:2016jqt,TheFermi-LAT:2017vmf,Cuoco:2017rxb,DiMauro:2019frs,DiMauro:2021raz,Cholis:2021rpp}).
Recently, Refs.~\cite{DiMauro:2021raz,Cholis:2021rpp} have provided comprehensive and updated results for the GCE properties.
They find that the GCE Spectrum Energy Distribution (SED) is peaked at a few GeV and has a high energy tail significant up to about 50 GeV.
The spatial distribution of the GCE is compatible with a DM template modeled with a generalized Navarro-Frenk-White density profile with slope $\gamma = 1.2\!-\!1.3$.
Moreover, the GCE centroid is located at the dynamical center of the Milky Way and its morphology is roughly spherically symmetric and does not change with energy.
Therefore, all the characteristics of the GCE are perfectly compatible with $\gamma$ rays produced from DM particles annihilating in the main halo of the Milky Way.
The GCE SED can be well modeled with DM particles of mass $(40\!-\!80)$ GeV  annihilating into $b\bar{b}$ with a thermal annihilation cross-section~\cite{Daylan:2014rsa,Calore:2014nla,Cuoco:2016jqt,Cuoco:2017rxb,DiMauro:2021qcf}, which is the correct cross-section to explain the observed relic density of DM in the Universe~\cite{Aghanim:2018eyx}.
Among several BSM theories that have been proposed to fit the GCE, the SHP model with a DM mass close to the Higgs resonance provides one of the simplest solutions (see, e.g.,~\cite{Okada:2013bna,Agrawal:2014oha,Mondal:2014goi,Duerr:2015bea,Cuoco:2016jqt,Cuoco:2017rxb}).

An alternative interpretation is associated to a population of millisecond pulsars located around the Galactic bulge that would produce a signal with properties compatible with the GCE (see, e.g.,~\cite{Bartels:2015aea,Lee:2015fea,Macias:2016nev,Bartels:2017vsx}).
Outbursts of cosmic rays (CRs) from the Galactic center have also been proposed as possible interpretations for the GCE (see, e.g.,~\cite{Carlson:2014cwa,Petrovic:2014uda,Gaggero:2015nsa}).

If DM is the actual origin of the GCE, $\gamma$ rays produced from these elusive particles could be detectable also from other astrophysical objects which are very dense of DM (see, e.g., for a review \cite{Fermi-LAT:2016afa}). Milky Way dwarf spheroidal galaxies (dSphs) are among the most promising targets for the indirect search of DM because gravitational observations indicate that they have a high DM density, i.e., a large mass-to-luminosity ratio of the order of $100-1000$ (see, e.g.,~\cite{Abdo_2010}). 
Since they do not contain many stars or gas, they have an environment with predicted low astrophysical backgrounds \cite{Geringer-Sameth:2018vjd}.
Analyses of known dSphs (see, e.g.,~\cite{Abdo_2010,Ackermann:2015zua,Lopez:2015uma,Fermi-LAT:2016uux,Calore:2018sdx,Hoof:2018hyn,2019MNRAS.482.3480P}) have imposed constraints on the DM interpretation of the GCE (see, e.g.,~\cite{Abdo_2010,Ackermann:2015zua,Lopez:2015uma,Fermi-LAT:2016uux,Calore:2018sdx,Hoof:2018hyn,2019MNRAS.482.3480P,DiMauro:2021qcf,DiMauro:2022hue}). However, combined analyses of dSph samples (see, e.g.,~\cite{DiMauro:2021qcf,DiMauro:2022hue}) or analyses of single objects (see, e.g.,~\cite{Hooper:2015ula}) have detected excesses at the level of $(1\!-\!3)\, \sigma$.

In addition to $\gamma$ rays, other cosmic particles, i.e.~messengers, could be produced from DM particle annihilation, such as antiprotons ($\bar{p}$).
Interestingly, different groups have found an excess of $\bar{p}$, with respect to the secondary production, in the data of AMS-02 \cite{PhysRevLett.117.091103} between $(5\!-\!20)$ GeV. 
Its significance has been found to vary from 1 to 5$\sigma$ depending on the analysis technique, CR propagation model and, more importantly, whether an estimation of the correlations in the AMS-02 systematic errors is considered \cite{Cuoco:2017rxb,Cui:2016ppb,Cuoco:2016eej,Cuoco:2019kuu,Cholis:2019ejx,Reinert:2017aga,Heisig:2020nse,DiMauro:2021qcf} (see, e.g.,~\cite{Heisig:2020jvs} for a review).
DM particles with a mass of $(60\!-\!80)$ GeV annihilating into $b\bar{b}$ can explain this excess, and a possible link to the GCE has been investigated~\cite{Cuoco:2017rxb,Cui:2016ppb,Cuoco:2019kuu,Cholis:2019ejx}.

Laboratories on Earth are also sensitive to BSM particles which could form DM\@. Experiments located at LHC, the largest operating particle collider to date, test BSM physics, including DM, making use of a variety of production processes and final state signals (see, e.g.,~\cite{Buchmueller:2017qhf} for a review). 
The LHC run 3 that started in July 2022 will further increase the sensitivity to DM signals.

Direct detection experiments such as LUX-ZEPLIN~\cite{LZ:2022ufs} (LZ) and XENONnT~\cite{XENON:2023sxq} are improving significantly our discovery potential of interactions between DM particles and detector atoms.
Both experiments have recently published the first results for the upper limits of the spin-independent cross sections for 60 days and 97.1 days of data taking, respectively, setting new limits by about a factor of two stronger with respect to previously existing measurements.
These experiments will push the limits closer to the neutrino fog \cite{Akerib:2022ort}.
Next generation direct detection experiments such as DARWIN \cite{DARWIN:2016hyl} will be able to investigate cross sections a factor of about 10 smaller than LZ and XENONnT (see, e.g.,~\cite{Schumann:2019eaa} for a review).

In this paper, we investigate the DM interpretation of the GCE 
within the SHP\@. 
We use a multi-messenger and multi-strategy approach. In particular, we perform a combined analysis of different cosmic messengers such as $\gamma$ rays, detected from dSphs and the Galactic center, and the flux data of $\bar{p}$ from AMS-02.
Then, we take advantage of the complementarity between direct and indirect detection and collider searches to verify whether the DM properties that explain the cosmic flux data are compatible with the LZ and LHC constraints. 
Finally, to find the model parameter space that matches all the observations, we compute the relic density, taking into account effects of kinetic decoupling  that are relevant for resonant DM annihilation~\cite{Binder:2017rgn,Binder:2021bmg} during freeze-out.

The paper is organized as follows. In Sec.~\ref{sec:model}, we introduce the SHP model and compute the annihilation cross sections and spectra relevant for indirect detection.
In Sec.~\ref{sec:RD}, we calculate the DM relic density. 
Constraints from collider and direct detection experiments are discussed in Secs.~\ref{sec:collider} and \ref{sec:DD}, respectively, before interpreting $\gamma$-ray and $\bar{p}$ data within the model  in Sec.~\ref{sec:ID}. Finally, in Sec.~\ref{sec:combined}, we combining direct, indirect, collider and cosmology data discussing implications for the viable parameter space. We conclude in Sec.~\ref{sec:conclusions}.

A \python package, called \singlet, enabling the fast calculation of the dark matter spectra, relic density, as well as direct detection and collider constraints within the model is available on \href{https://github.com/dimauromattia/SingletScalar_DM}{\github}.

%
%
%%%%%%%%%%%%%%%%%%%%%%%%%%%%%%%%%%%%%%%%%%%%%%%%%%%%%%%%
\section{Singlet scalar dark matter with a Higgs portal}
\label{sec:model}

\subsection{Overview of the model}
\label{sec:overview}

The SHP model is among the simplest DM models extending the SM by just a DM particle candidate, $S$, taken to be a real scalar and a singlet under the SM gauge group. The Higgs portal provides a unique gauge invariant and renormalizable interaction that couples $S$ to the SM without the need of additional mediator particles. The model has first been considered in Ref.~\cite{SILVEIRA1985136} and has since been extensively studied in the literature, see, e.g., Refs.~\cite{McDonald:1993ex,Burgess:2000yq,Davoudiasl:2004be,Ham:2004cf,OConnell:2006rsp,Profumo:2007wc,Barger:2007im,Lerner:2009xg}.

The BSM part of the model Lagrangian reads    
\begin{equation}
\Delta\mathcal{L}_{\mathrm{SHP}}=
\frac{1}{2} \partial_\mu S \partial^\mu S-\frac{1}{2} \mu_S^2 S^2-\frac{1}{4}\lambda_S S^4-\frac{1}{2}\lambda_{HS} H^{\dagger} H S^2,
\label{eq:SHP1}
\end{equation}
where $H$ is the SM Higgs doublet and $\mu_S,\lambda_S$, and $\lambda_{HS}$ are free parameters of the theory.
In Eq.~\eqref{eq:SHP1}, the terms refer to (from left to right): 
the $S$ kinetic term, the bare $S$ mass, the $S$ quartic self-coupling and the Higgs-portal coupling. 
The Lagrangian respects a discrete $Z_2$ symmetry that guarantees the stability of $S$. Under this symmetry all SM particles are assumed to be even while $S$ is odd ($S \rightarrow -S$).

After spontaneous electroweak symmetry breaking (EWSB), the Higgs boson acquires a vacuum expectation value, $\langle H \rangle  = v_0/\sqrt{2}$,  with $v_0 = 246.2$ GeV, and  
Eq.~\eqref{eq:SHP1} becomes:
\begin{multline}
\Delta\mathcal{L}_{\mathrm{SHP}}=\frac{1}{2} \partial_\mu S \partial^\mu S - \frac{1}{2} m_S^2 S^2-\frac{1}{4}\lambda_S S^4\\
-\frac{1}{4}\lambda_{HS} h^2 S^2-\frac{1}{2} v_0 \lambda_{HS} h S^2,
\label{eq:SHP2}
\end{multline}
where $h$ is the Higgs field and $m^2_S = \mu_s^2 + 1/2\, v_0^2 \lambda_{HS}$ is the mass of $S$ in the broken phase.
\begin{figure*}[t]
\includegraphics[scale=0.8]{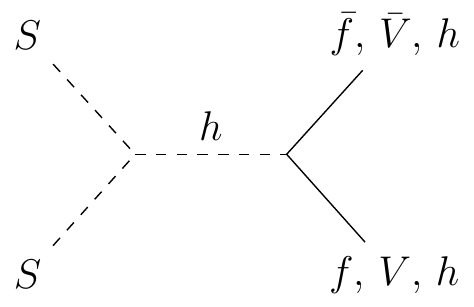}\hspace{1cm}
\includegraphics[scale=0.8]{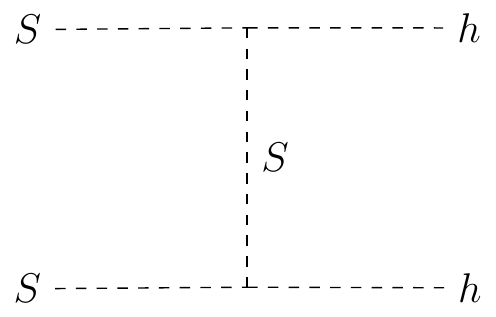}\hspace{1cm} 
\includegraphics[scale=0.8]{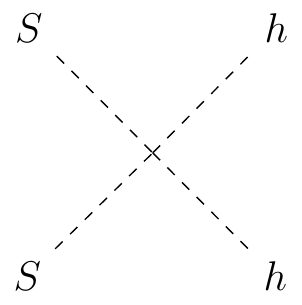}
\caption{Tree-level annihilation Feynman diagrams of the SHP model. The symbol $f$ refers to charged fermions, while the $V$ refers to either $W^{\pm}$, $Z$. The central diagram includes also the u-channel.} 
\label{fig:shp_diagrams_1}
\end{figure*}
\begin{figure*}[t]
\includegraphics[scale=0.8]{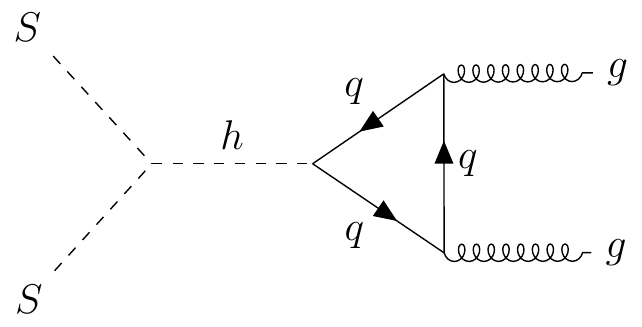}
\caption{Feynman diagram for the loop-induced annihilation $SS \to gg$. The symbol $q$ stands for quarks.} 
\label{fig:shp_diagrams_2}
\end{figure*}

The quartic scalar self coupling term, proportional to $\lambda_S$, is of importance for the stability of the electroweak vacuum and the perturbativity of the model~\cite{Gonderinger:2009jp} but does not affect the DM phenomenology.
The latter is solely governed by the last two interaction terms in Eq.~\eqref{eq:SHP2}. 
Thanks to these terms, $S$ can annihilate into all SM particles through the Higgs portal with a coupling that is proportional to $\lambda_{HS}$.
The annihilation process is relevant for thermalization and freeze-out of $S$ in the early Universe and can lead to the  production of SM particles in our Galaxy today. Moreover, the Higgs boson may decay into the scalar $S$ producing an observable signature at colliders. Finally, the scattering of $S$ off quarks, mediated by $h$, could produce recoil events in direct detection experiments.  

The SHP model is very simple, with only two parameters that are relevant for the DM physics: the physical DM mass $m_S$ and the Higgs portal coupling $\lambda_{HS}$.
This allows us to straight-forwardly constrain the parameters of the model through the relic density constraint and from direct and collider searches and astrophysical observations.
In Figs.~\ref{fig:shp_diagrams_1} and~\ref{fig:shp_diagrams_2}, we show the Feynman diagrams relevant for the annihilation process within the SHP model.
The model contains one diagram for each fermion, one for each gauge boson $W^{\pm}$ and $Z$ and three diagrams for the annihilation into the Higgs boson.
We do not show the channels with the production of $\gamma \gamma$ and $\gamma Z$ or $\gamma h$ that are loop induced, thus providing a subdominant contribution to the cross section.

\subsection{Annihilation cross-section}
\label{sec:sigmav}

Except for the channel with $h h$ final states, the cross section $\sigma$ multiplied by the M\"oller velocity $\vMol$ of DM pairs annihilating into SM particles $i$ can be expresses as \cite{Cline:2013gha}:
\begin{equation}
\label{eq:annisigma}
\sigma \vMol=\frac{2 \lambda_{HS}^2 v_0^2}{\sqrt{s}}\left|D_h(s)\right|^2 \Gamma_{h \rightarrow i}(\sqrt{s}),
\end{equation}
where $\Gamma_{h \rightarrow i}(\sqrt{s})$ is the partial decay width into state $i$ of the SM Higgs boson evaluated at energy $\sqrt{s}$. $D_h(s)$ is the Higgs propagator defined as:
\begin{equation}
\label{eq:propagator}
\left|D_h(s)\right|^2=\frac{1}{\left(s-m_h^2\right)^2+m_h^2 \Gamma_h^2} \,,
\end{equation}
where $\Gamma_h$ is the total Higgs width, which includes all the kinematically allowed partial decay widths, as well as the invisible Higgs width $\Gamma_{h,\text{inv}}$. For the former, we adopt the theoretical prediction from Ref.~\cite{LHCHiggsCrossSectionWorkingGroup:2011wcg}, $\Gamma_{h, \text{SM}}=4.1$ MeV. 
For the latter, we employ the tree level result, given by:
\begin{equation}
\Gamma_{h,\text{inv}}=\frac{\lambda_{HS}^2 v_0^2}{32 \pi m_h}\left(1-\frac{4 m_S^2 }{m_h^2} \right)^{1 / 2},
\label{eq:Gh}
\end{equation}
where $m_h$ is the Higgs mass. We use $m_h =125$ GeV as recently measured by ATLAS \cite{ATLAS:2022net}. 
The expression for $\sigma \vMol$ written in Eq.~\eqref{eq:annisigma} is particularly convenient because very precise theoretical calculations and measurements are present for the quantity $\Gamma_{h\rightarrow i}$ (see, e.g.,~\cite{LHCHiggsCrossSectionWorkingGroup:2011wcg}).
\begin{figure*}[t]
\includegraphics[width=0.49\linewidth]{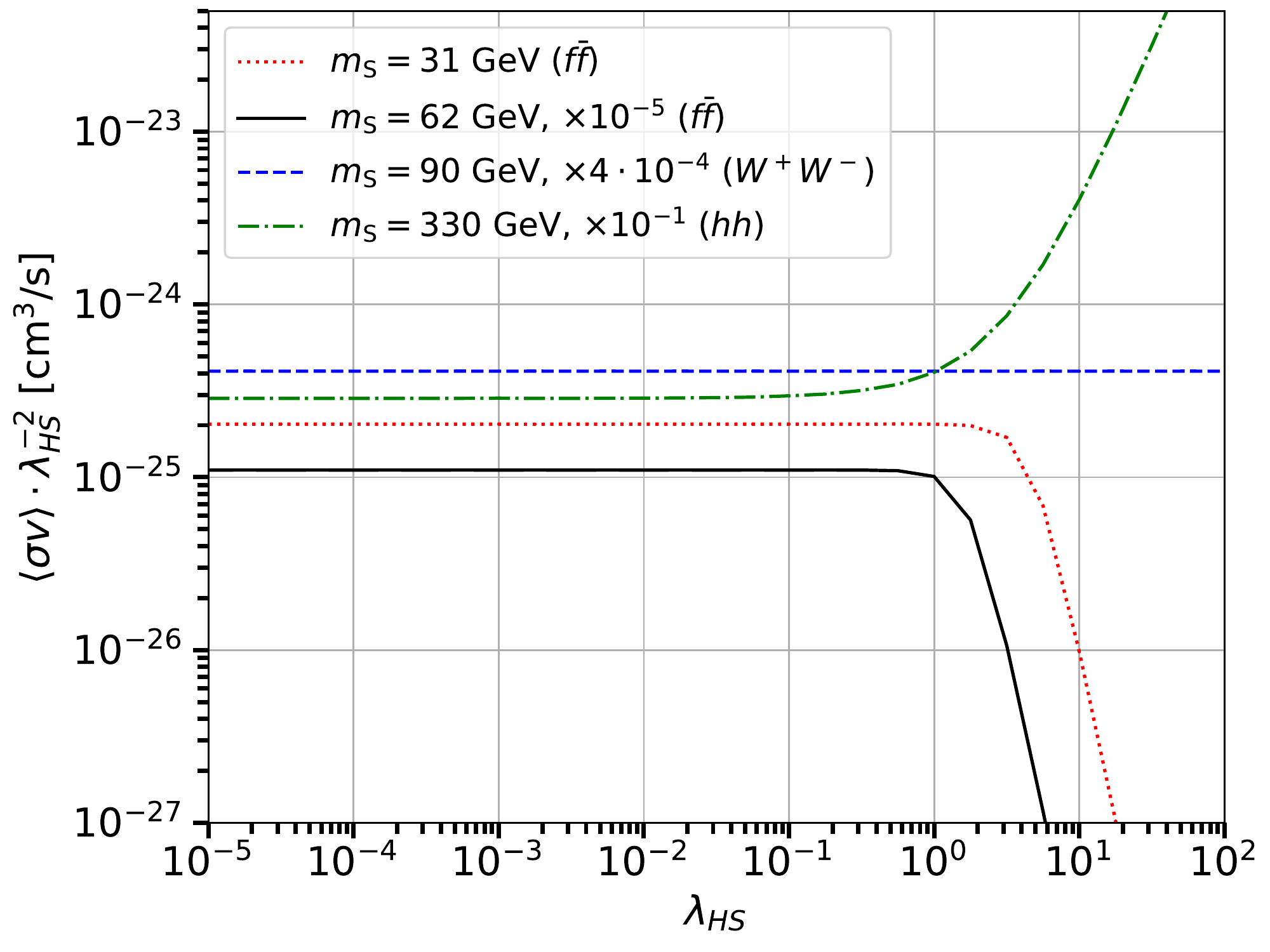}
\includegraphics[width=0.49\linewidth]{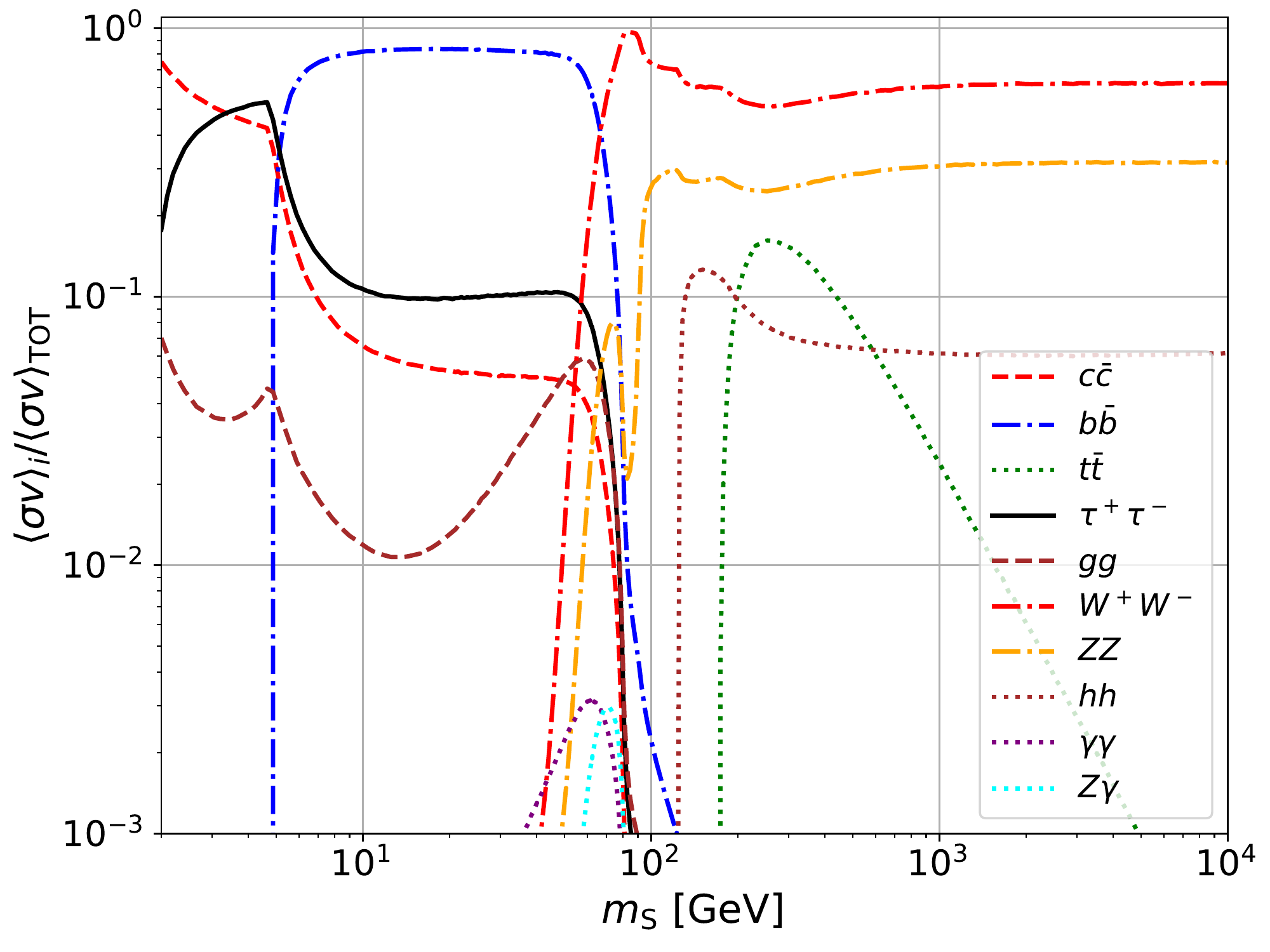}
\caption{Left panel: Annihilation cross-section rescaled by $\lambda_{HS}^{-2}$ as a function of the coupling parameter $\lambda_{HS}$ for four different values of $m_S$.
For $m_S=31$ GeV and $m_S=62$ GeV, \sigmav is dominated by the $f\bar{f}$ final state;  \sigmav for $m_S=90$ GeV is dominated by the $W^{\pm}$ channel while \sigmav for $m_S=330$ GeV takes most of its contribution from the Higgs production.
Right panel: Ratio between the annihilation cross-section for the different channels ($\langle \sigma v \rangle_i$) and the total one ($\langle \sigma v \rangle_{\rm{tot}}$) as a function of the DM mass, computed for a fixed $\lambda_{HS} = 0.01$.
}
\label{fig:annih}
\end{figure*}

In Eq.~\eqref{eq:propagator}, we write down the propagator within the commonly used fixed-width prescription. This is typically a good approximation except when $\Gamma_{h}$ is a rapidly varying function of $s$. This can, in particular, happen in the resonant region, $m_S\approx m_h/2$, where the invisible decay channel opens up close to the resonance~\cite{Heisig:2019vcj}.
In this case, the running of the Higgs width has to be taken into account by replacing $m_{h}$ in Eq.~\eqref{eq:Gh} with the center-of-mass energy $\sqrt{s}$ of the process. However, this effect is only relevant for sizeable coupling, as quantified at the end of Sec.~\ref{sec:RD}.

For the indirect detection signals, the cross section is averaged over the DM particle velocity distribution, for which we adopt the Standard Halo Model (see, e.g.,~\cite{Evans:2018bqy}). However, the cross sections in the SHP are essentially s-wave, i.e.~they do not depend on the velocity. The only region of the parameter space where the velocity-averaged annihilation cross-section $\left\langle\sigma \vMol\right\rangle$ depends on $v$ is very close to the Higgs resonance. While this region is of high importance to our analysis, we nevertheless find that the velocity-dependence is relevant only for velocities significantly larger than the typical DM velocity in the Galactic halo. Therefore, for the typical Galactic velocities, which are of the order of $10^{-3}c$, the velocity-dependent term at the resonance contributes only up to 1\%, while being negligible otherwise. 

The expression of the annihilation cross-section for all possible annihilation channels is provided in, e.g., Ref.~\cite{Duerr:2015aka}. 
Here, we simply summarize the formulae for the relevant leading-order cross-sections and discuss their dependence on the parameters $m_S$ and $\lambda_{HS}$. 
The scaling of the velocity averaged annihilation cross-section for different DM masses is shown in Fig.~\ref{fig:annih} and it will be used to guide the discussion.
The annihilation cross-section into a pair of fermions is given by (see, e.g.,~\cite{Duerr:2015aka}):
\begin{equation}
\sigma = \frac{\lambda_{HS}^2 m_f^2 N_c (s- 4m_f^2)^{3/2}}{8\pi s \sqrt{s-4m_s^2}[(s-m_h^2)^2+\Gamma_h^2 m_h^2]},
\label{eq:csf}
\end{equation}
where $N_c = 3$ for quarks and $N_c = 1$ for leptons, and $m_f$ is the fermion mass.
If the $S$ mass is above the Higgs resonance ($m_S>m_h/2$), $\Gamma_h$ is independent from $\lambda_{HS}$ and the annihilation cross-section into fermions scales as $\lambda_{HS}^2$ (see Fig.~\ref{fig:annih} for $m_S=90$ GeV). This holds also for $\lambda_{HS}<0.01$ when $m_S < m_h/2$, because the decay of the Higgs boson into DM particles is negligible with respect to the decay into SM particles, and $\Gamma_h$ remains approximately equal to the SM one. Instead, for $m_S<m_h/2$ and $\lambda_{HS}>0.01$, $\Gamma_h$ gets a contribution which increases proportional to $\lambda_{HS}^2$ [see Eq.~\eqref{eq:Gh}], thus adding an additional dependence on $\lambda_{HS}$ in the cross section, so that $\sigma \propto \lambda_{HS}^2/(1 + k \lambda_{HS}^4)$.
Note, however, that the second term in parentheses is subleading unless $\lambda_{HS}$ approaches the non-perturbative regime in which the validity of the expression becomes questionable.

The cross section into $W^{\pm}$ and $Z$ gauge bosons is given by:
\begin{equation}
\sigma = \frac{\lambda_{HS}^2 \sqrt{s- 4m_V^2}(s^2-4 m_{V}^2s + 12 m_{V}^4)}{16\pi s \sqrt{s-4m_S^2}[(s-m_h^2)^2+\Gamma_h^2 m_h^2]},
\label{eq:csV}
\end{equation}
where $V$ stands for $W^{\pm}$ and $Z$.
Most of the contribution from the annihilation into $W^{\pm}$ and $Z$ comes from DM masses above the resonance for which $\Gamma_h$ does not depend on $\lambda_{HS}$. Therefore, the annihilation cross-section into $W^{\pm}$ and $Z$ scales as $\lambda_{HS}^{2}$ in the most relevant parameter space (see Fig.~\ref{fig:annih} for $m_S=90$ GeV that is dominated by the $W^{\pm}$ channel).

The cross section for annihilation into a pair of Higgs bosons gives rise to a more lengthy expression that we omit. However, in the non-relativistic limit, $s\rightarrow 4 m_S^2$, it reduces to:
\begin{equation}
\sigma v_{\rm{rel}} = \frac{\lambda_{HS}^2 \sqrt{1-\frac{m_h^2}{m_S^2}}[m_h^4-4 m_S^4 +2 \lambda_{HS} v^2 (4 m_S^2-m_h^2)]^2}{64\pi m_S^2 (m_h^4-6m_h^2m_S^2+8m_S^4)^2}.
\label{eq:csH}
\end{equation}
In this case, for $\lambda_{HS}<0.1$ the annihilation cross-section scales as $\lambda_{HS}^2$. In contrast, for larger couplings, the polynomial term in the numerator becomes more relevant than all of the other terms, and the cross section scales as $\lambda_{HS}^4$. 
This is shown in Fig.~\ref{fig:annih} for the case of $m_S=330$ GeV, that  is dominated by the annihilation into the Higgs boson.

We also take into account the annihilation into the gluon final states $gg$ (see Fig.~\ref{fig:shp_diagrams_2}). This is a loop-induced process, which is typically suppressed with respect to the tree-level annihilation channels. In fact, its contribution is at most $7\%$ for $m_S\approx 60$ GeV. We do not take into account the annihilation channels into $\gamma \gamma$, $\gamma Z$ and $\gamma h$ because their contribution is at most at the few per mille level. 

While we have explicitly shown here some expressions for the annihilation cross-section, useful to guide the discussion, we nevertheless calculate \sigmav for all the annihilation channels using \maddm \footnote{\url{https://launchpad.net/maddm}} \cite{Backovic:2013dpa,Backovic:2015cra,Ambrogi:2018jqj,Arina:2021gfn}, a plugin of \mg~\cite{Alwall:2014hca,Frederix:2018nkq}. We utilise the UFO implementation~\cite{Degrande:2011ua,Darme:2023jdn} for the SHP model. \maddm is linked to \pythia for hadronization and showering. In this work, we use \maddm version 3.2 \cite{Arina:2021gfn} on top of \mg version 2.9.9 (LTS) and \pythia version 8.3 \cite{Bierlich:2022pfr}.

For the annihilation channels into massive gauge or Higgs bosons, we include their off-shell contributions.
This is particularly relevant for $W^{\pm}$ and $Z$ final states, while it has very minor effect in the case of the $h$ bosons.
To this end, we calculate the cross section for the four-body diagrams, where two bosons $V V^\prime$ are produced off-shell from DM annihilation, subsequently decaying into four fermions $f$:
\begin{equation}
S S \rightarrow h \rightarrow V V^\prime \rightarrow 4f.
\label{eq:4bodydiagram}
\end{equation}

The relative contributions of the different channels, $\langle \sigma \vMol \rangle_i/  \langle \sigma \vMol \rangle_{\rm{tot}}$, is shown in Fig.~\ref{fig:annih} as a function of the DM mass, with $\lambda_{HS}=0.01$.
For very small $S$ masses, the annihilation happens predominantly into $c\bar{c}$ quarks and $\tau^+ \tau^-$ leptons. For $m_S = (5-50)$ GeV instead, the main contribution comes from $b\bar{b}$ annihilation. Finally, for large DM masses the main channels are the one with gauge and Higgs bosons.
The effect of the off-shell contribution of the gauge bosons is particularly relevant for the annihilation involving $W^{\pm}$ bosons, which becomes the main channel even below the on-shell production threshold $m_S < m_W$.

For DM masses above $5$ GeV, the contribution of $\tau^+ \tau^-$ annihilation is larger than $c\bar{c}$.
This may seem to be counter-intuitive since, from Eq.~\eqref{eq:csf}, the cross sections for the two channels are $\sigma_{c \bar{c}} \propto 3 m_c^2$ and $\sigma_{\tau^+ \tau^-} \propto m_\tau^2$.
Accordingly, one may expect $\sigma_{c \bar{c}}/ \sigma_{\tau^+ \tau^-} = 3m_c^2/m_{\tau}^2>1$.
However, taking into account the running quark masses in our calculation, we find $m_c \leq 1$ GeV \cite{Gizhko:2017fiu} for $m_S \geq 3$ GeV, which leads to $\sigma_{c \bar{c}}< \sigma_{\tau^+ \tau^-}$.

\subsection{Source energy spectra}
\label{sec:spectra}
\begin{figure*}[t]
\includegraphics[width=0.49\linewidth]{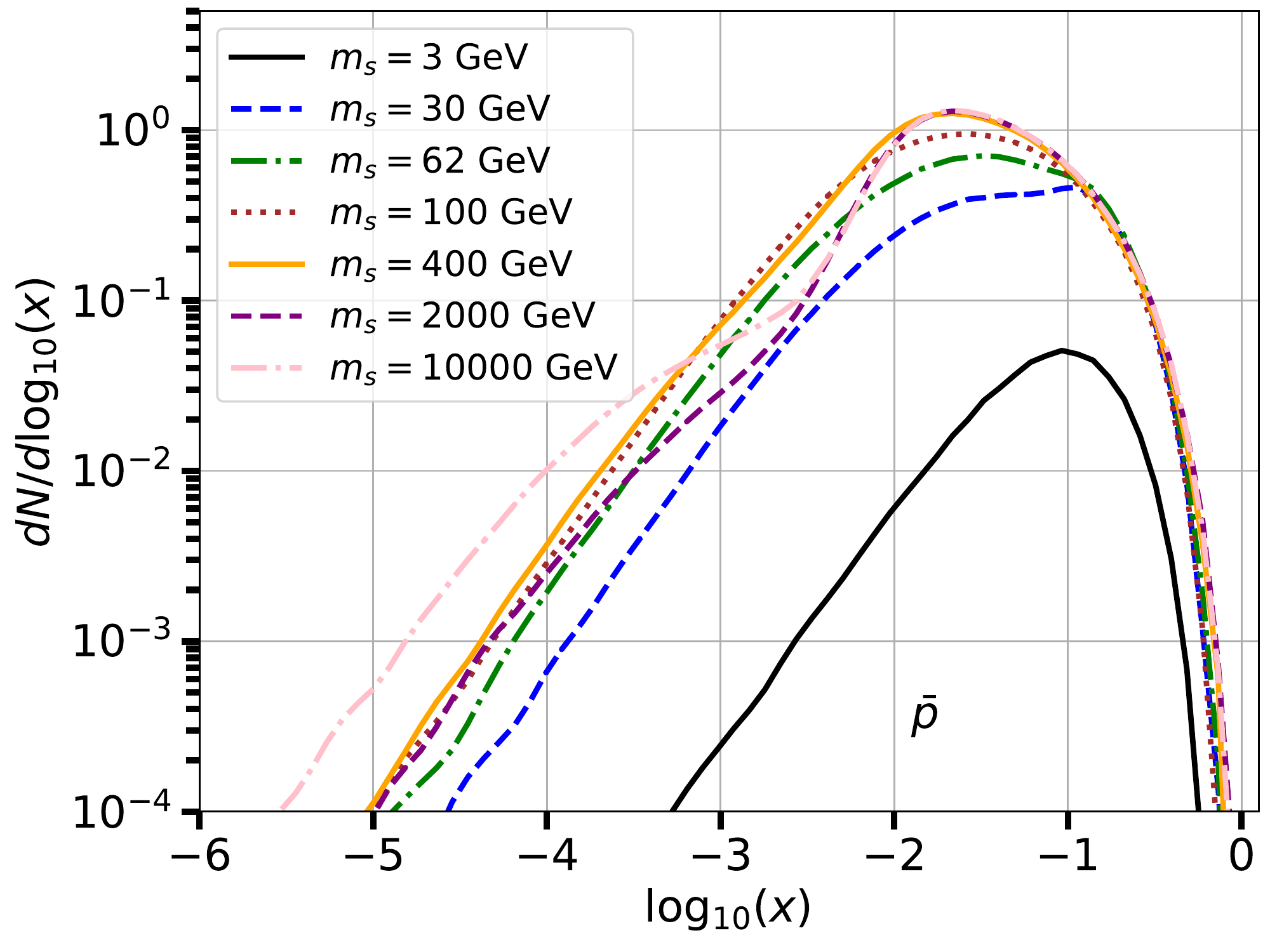}
\includegraphics[width=0.49\linewidth]{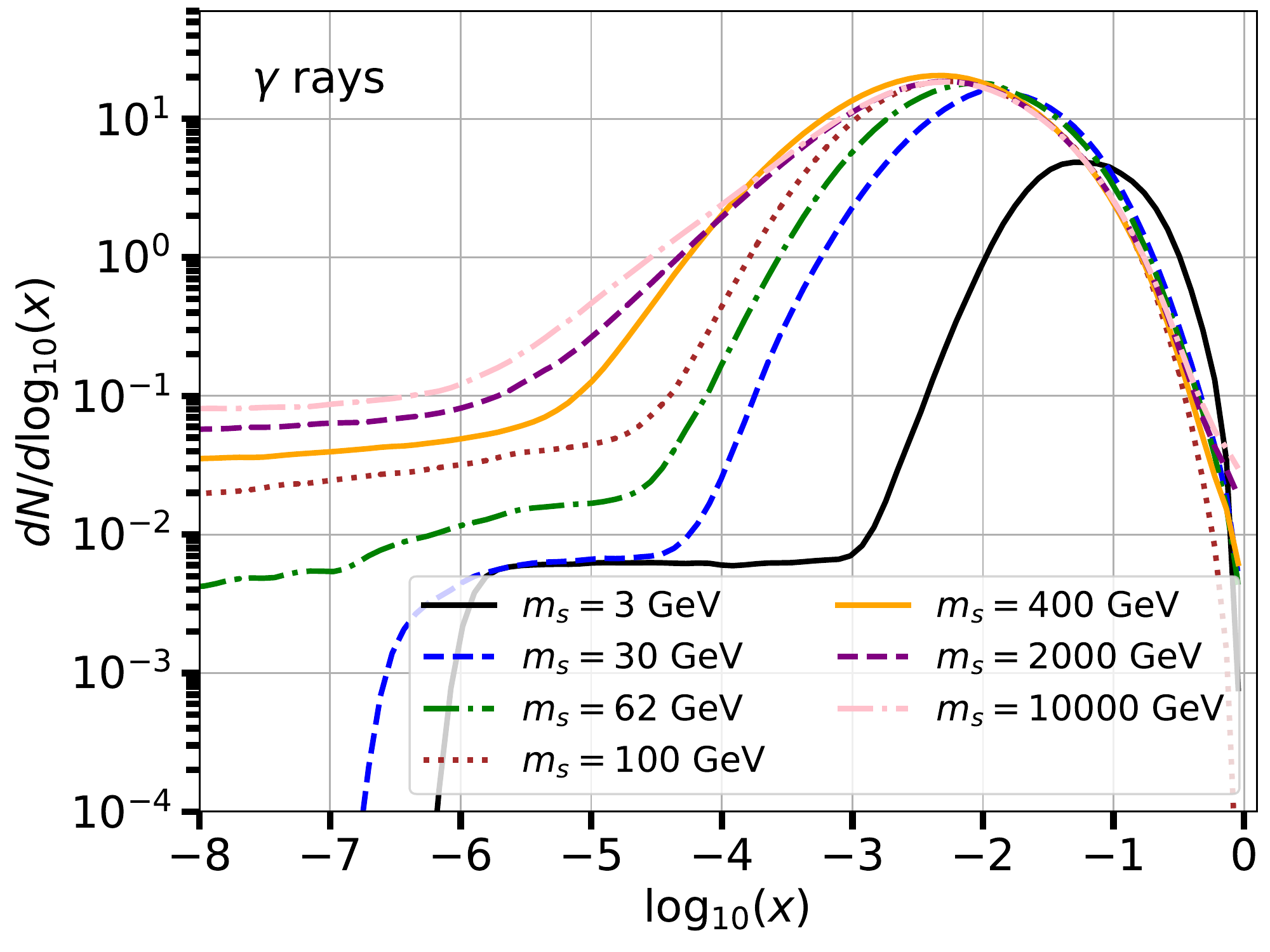}
\caption{Left panel: Energy spectra ($\odv{N}/{\log_{10}(x)}$) of $\bar{p}$ for the SHP model plotted as a function of $x=K/m_S$, where $K$ is the kinetic energy. We show the energy spectra for different DM mass values ranging from 3 GeV to 10 TeV. Right panel: Same as left for the $\gamma$ rays.}
\label{fig:DMspectra}
\end{figure*}
The energy spectra $\odv{N_p}/{E}$ of particles such as photons and antiprotons, produced from DM annihilation is an important ingredient for the indirect detection of DM (see Sec.~\ref{sec:ID}).
In order to calculate these spectra we use the code \maddm, which first evaluates the annihilation cross-sections. Subsequently, \maddm calls \pythia to generate the spectra for the different annihilation channels, and  determines the final energy spectra by averaging the results obtained for each channel by $\langle \sigma v \rangle_i/\langle \sigma v \rangle_\text{tot}$.

We include in the computation all channels (tree level and loop induced) that we consider for the calculation of the annihilation cross-sections (see Sec.~\ref{sec:sigmav}).
The spectra we generate take into account final-state-radiation (FSR) and electroweak emission of weak gauge bosons off fermions, which are included by \pythia. With respect to the computation of the tables provided in Ref.~\cite{Cirelli:2010xx} there are two important additions. First, we include the contribution of off-shell $W^{\pm}$ and $Z$, which could be very relevant between 60--90 GeV, as seen in Fig.~\ref{fig:annih}.
Second, we take into account the polarization of $W^{\pm}$ and $Z$ bosons.  This effect is particularly relevant for the spectra of antiprotons. For all the technical details about FSR, electroweak corrections and inclusion of $W^{\pm}$ and $Z$ polarization see Appendix \ref{app:spectrum}.

We compute the DM spectra for masses between 2 GeV and 20 TeV generating $10^6$ events per point. 
For DM masses above $110$ GeV (where the $hh$ channel introduces a dependence of the relative contributions of to annihilation on $\lambda_{SH}$) we vary $\lambda_{SH}$ between $10^{-2}$ and $10$.\footnote{We remind the reader that we are considering also the contribution of off-shell Higgs bosons. Therefore, the annihilation channel into $hh$ is present also below $m_h=125$ GeV.} 
In Fig.~\ref{fig:DMspectra}, we show the results for the DM spectra of $\gamma$ rays and $\bar{p}$ for DM masses between 3 GeV and 10 TeV.
The energy spectra, reported in terms of $\odv{N}/{\log_{10}(x)}$, where $x=K/m_S$ with $K$ being the kinetic energy, have a peak at an energy value that shifts towards lower values when increasing the DM mass.
The $\bar{p}$ spectrum is decreasing for $\log_{10}(x)<-2$.
Instead, the $\gamma$-ray spectrum has a flattening at around $\log_{10}(x)<[-6,-3]$ depending on the mass. This change of shape is due to the contribution of the electroweak corrections such as FSR.

\section{Relic density}
\label{sec:RD}

The DM average density has recently been measured by the Planck experiment to be $\Omega_{\rm{DM}} h^2 = 0.120$ with an uncertainty at the level of $1\%$ \cite{Planck:2018vyg}.
In general, the theoretical calculation of the DM relic abundance requires to solve the Boltzmann equations for the DM phase-space density $f_S(\vec{p})$ in an expanding Friedmann-Robertson-Lema\^itre-Walker Universe~\cite{Kolb:1990vq,Edsjo:1997bg}:
\begin{equation}
E(\partial_t - H \vec{p} \cdot \nabla_{\vec{p}}) f_{S} = \mathcal{C} \left[f_S\right]\,,
\label{eq:RD1}
\end{equation}
where $E$ and $\vec{p}$ are the energy and momentum of the $S$ particle, and $H$ is the Hubble expansion rate.
The term $\mathcal{C}$ is the collision operator that takes into account all the interactions between DM and SM particles.
In particular, $\mathcal{C}$ contains the operator for elastic scattering ($\mathcal{C}_{\rm{el}}$) and annihilation ($\mathcal{C}_{\rm{ann}}$). 
Elastic collisions are responsible for maintaining the kinetic equilibrium while inelastic collisions keep the chemical equilibrium.
See Ref.~\cite{Binder:2017rgn} for a complete description of $\mathcal{C}_{\rm{el}}$ and $\mathcal{C}_{\rm{ann}}$.

For the canonical freeze-out of weakly interacting massive particles, a number of approximating assumptions are usually applied that significantly simplify Eq.~\eqref{eq:RD1} and its numerical solution.
Most significantly, this concerns the assumption of kinetic equilibrium between DM and the SM bath throughout the entire process of chemical decoupling. The assumption that the distribution of $S$ particles remains proportional to the thermal one, $f_S\propto f_{S,\text{eq}}$, simplifies the partial differential equation Eq.~\eqref{eq:RD1} into an ordinary differential equation for the DM number density, i.e.,~the well-known Zeldovich-Okun-Pikelner-Lee-Weinberg equation \cite{YaBZel'dovich_1966,PhysRevLett.39.165}:
\begin{equation}
\frac{\mathrm{d} n_S}{\mathrm{~d} t}+3 H n_S=-\left\langle\sigma \vMol\right\rangle_T \left(n_S^2-n_{S, \mathrm{eq}}^2\right),
\label{eq:RD2}
\end{equation}
where $n_S=g_\chi \int d^3 p /(2 \pi)^3 f_S(p)$ and $\left\langle\sigma \vMol\right\rangle_T$ is the thermally averaged cross-section at temperature $T$. In the non-relativistic regime, the $S$ particle phase-space density follows Maxwell-Boltzmann statistics $f_{S,\text{eq}}\simeq \exp{(-E(p)/T)}$ and $\left\langle\sigma \vMol\right\rangle_T$ reads:
\begin{equation}
\left\langle\sigma \vMol\right\rangle_T=\int_{4 m_S^2}^{\infty} \mathrm{d} s \frac{s \sqrt{s-4 m_S^2} K_1(\sqrt{s} / T) \sigma \vMol}{16 \, T m_S^4 K_2^2\left(m_S / T\right)}\,,
\end{equation}
where $K_i$ are the modified Bessel functions of order $i$.
The assumption of kinetic equilibrium is often well justified as elastic scattering processes can easily be orders of magnitude larger than the annihilation processes that initiate chemical equilibrium. This is due to the large number density of light SM particles DM can scatter off (compared to the Boltzmann suppressed number density of heavy DM particles during freeze out). However, this reasoning does not carry over to annihilation via a resonant Higgs in $s$-channel. First, elastic scattering processes are not resonantly enhanced and, hence, suppressed compared to annihilation. Second, as the Higgs-SM couplings are proportional to the SM particles masses, coupling to light particles -- with unsuppressed number densities -- are small. As a result, kinetic equilibrium can break down during chemical freeze-out and the above assumption is unjustified~\cite{Binder:2017rgn}.
In this case, the full version of the Boltzmann equation must be solved, with the inclusion of the elastic collision term.
\begin{figure*}[t]
\includegraphics[width=0.49\linewidth]{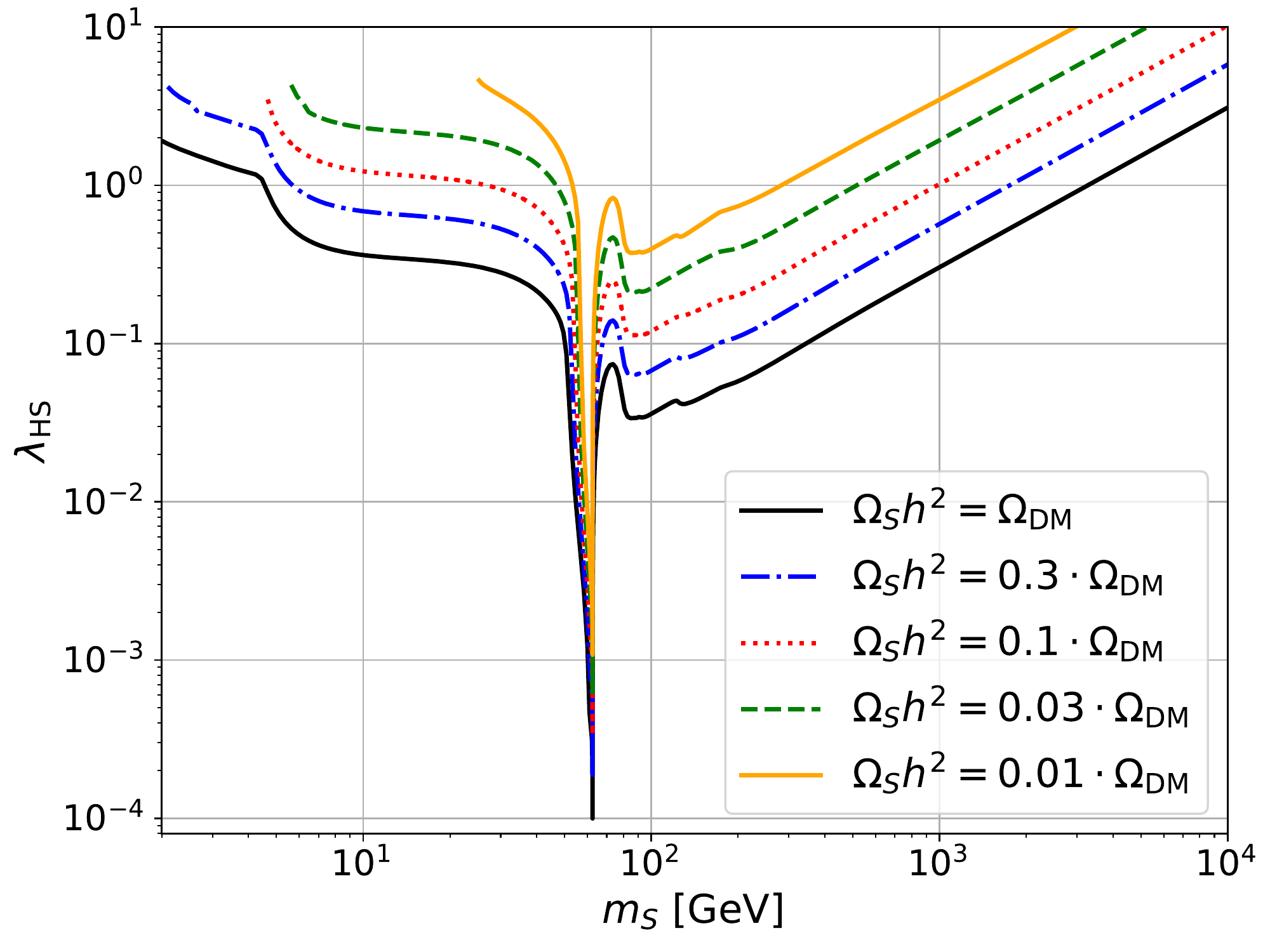}
\includegraphics[width=0.49\linewidth]{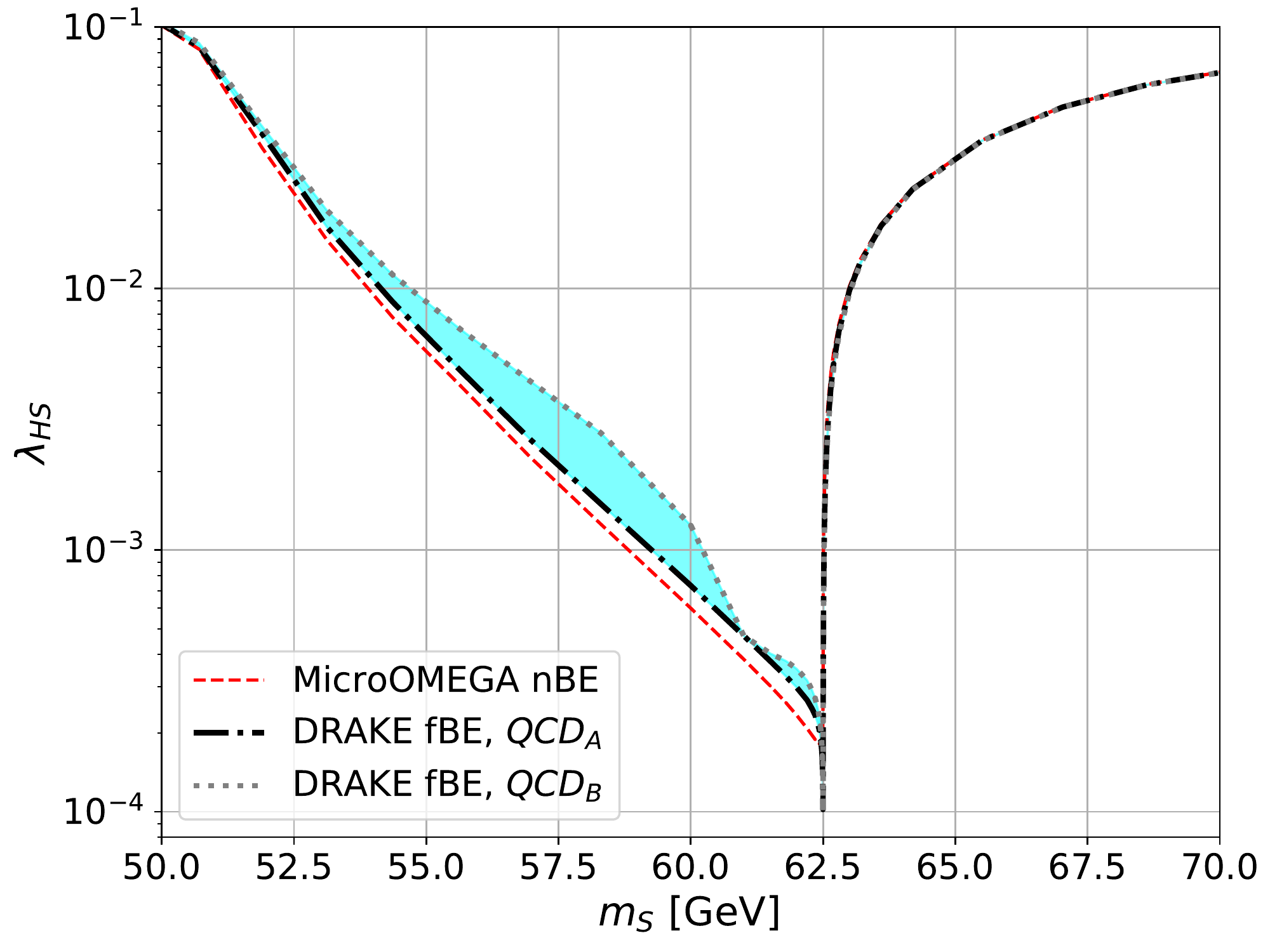}
\caption{Left panel: Combination of parameters $\lambda_{HS}$ and $m_S$ reproducing $\Omega_S h^2$  within a fraction (from $1\%$ to $100\%$) of the measured value of the relic abundance obtained by the Planck experiment, $\Omega_{\rm DM} h^2$. Here $\Omega_S h^2$ is computed solving the full Boltzmann equation taking into account both the elastic and annihilation collision operators (model {\tt fBE} with QCD$_B$ with \drake).
Right panel: Comparison of the results obtained by solving the Boltzmann equation with both \micromegas and \drake [see Eq.~\eqref{eq:RD1}]. The \micromegas line  accounts only for the annihilation processes ({\tt nBE}); the \drake lines assume always  {\tt fBE} for the two extreme QCD phase transition choices, {\tt QCD}$_A$ and {\tt QCD}$_B$.}
\label{fig:RD1}
\end{figure*}

We compute the relic density for $S$, hereafter called $\Omega_S h^2$, in both setups, labelled as:
\begin{itemize}
\item {\tt fBE}: the full solution of Eq.~\eqref{eq:RD1};
\item {\tt nBE}: the solution of Eq.~\eqref{eq:RD2}, i.e.,~assuming kinetic equilibrium during freeze-out.
\end{itemize}
In Ref.~\cite{Binder:2017rgn} the authors found that the relic abundance predicted following the approximated approach {\tt nBE} can differ by up to an order of magnitude from the calculation obtained with the full Boltzmann equation, {\tt fBE}.
We performed the relic density calculation using the code \drake~\cite{Binder:2021bmg}, developed by the authors of Ref.~\cite{Binder:2017rgn}.
This code solves the Boltzmann equation with both the models {\tt nBE} and {\tt fBE}.
For a detailed explanation, see Ref.~\cite{Binder:2021bmg}. 

For the values of $m_S$ we considered, the freeze-out takes place at temperatures around a few GeV\@.
That means we are close to the region in which the QCD phase transition takes place, after which quarks become confined in baryons and mesons.
To take into account this effect, we considered two different physics models, which are implemented in the \drake code and studied in Ref.~\cite{Binder:2017rgn}.
These two models are extreme scenarios that are intended to bracket the uncertainties:
The first model, named {\tt QCD}$_{A}$, represents the case maximizing the elastic scattering for which all quarks are free and present in the plasma (according to their equilibrium abundance) down to temperatures of $T_c = 154$ MeV.
The second model, named ({\tt QCD}$_{B}$), minimizes the elastic scattering: only light quarks (u, d, s) can contribute, and only considering temperatures above $4T_c \approx 600$ MeV.
This threshold ensures that hadronization effects are negligible.

For comparison, we also compute $\Omega_S h^2$ in the setup \texttt{nBE} with \micromegas\footnote{\url{https://lapth.cnrs.fr/micromegas/}}~\cite{Belanger:2006is,Belanger:2013oya,Barducci:2016pcb,Belanger:2018ccd}.
We consider the SHP model implementation shipped with the code.
Both \drake and \micromegas take into account the off-shell contributions of $W^{\pm}$ and $Z$ gauge bosons, and the contribution coming from loop-induced annihilation into gluons.

The left panel of Fig.~\ref{fig:RD1} shows the combination of parameters $\lambda_{HS}$ and $m_S$ for which the relic density $\Omega_{S}h^2$ corresponds to the measured value $\Omega_{\rm DM} h^2 = 0.120$.
We show the computation of the {\tt fBE} model in the {\tt QCD}$_B$ scenario with \drake.
The first consideration we can do regards the resonance region: for $m_S\approx m_h/2$ the value of $\lambda_{HS}$ decreases very quickly.
This can be explained with the resonant enhancement of the annihilation cross-section for this particular parameter choice.
As a result, the correct value of the relic abundance necessarily requires a very small value of the coupling $\lambda_{HS}$.
The right panel of Fig.~\ref{fig:RD1} shows the parameter $\lambda_{HS}$ accounting for the correct relic density value as a function of $m_S$ in the resonance region.
We show the computations performed with \micromegas (which includes only the {\tt nBE} model) and with \drake (for both the {\tt fBE} {\tt QCD}$_A$ and {\tt QCD}$_B$ models).
The difference between the {\tt fBE} case and the normal approach {\tt nBE} varies within a  factor 0.5 and 2, and it is present in the range of DM masses between $(50-67)$ GeV.
In particular, around 57 GeV, the {\tt fBE} result is approximately a factor of 2 larger than the {\tt nBE} one, while on the Higgs pole it differs by a factor of $\sim 0.5$.
In the figure we also show the difference between the {\tt QCD}$_A$ and {\tt QCD}$_B$ assumptions, which is maximal in the range $(55-60)$ GeV, where it is at the level of $40-50\%$.
This region represents the uncertainty due to the QCD phase transition.
This is our main result for the computation of the relic density, which properly takes into account the kinetic and thermal parts of the Boltzmann equation, thanks to the implementation of the {\tt fBE} model, additionally accounting for the QCD phase transition effect in the resonance region.

There are further corrections to the annihilation channel that can potentially become important close to the resonance. As we will discuss in Sec.~\ref{sec:collider} in the context of LHC constraints, for $m_S$ very close to $m_h/2$, the centre-of-mass energies relevant in the annihilation process are around the $SS$ threshold. Accordingly, the change of the Higgs width due to the opening of the $SS$ decay channel can become relevant and the fixed-width calculation can break down calling out for a running-width description~\cite{Heisig:2019vcj}. However, for annihilation, we find that this effect is larger than a few percent only for $\lambda_{HS} > 0.3$ in the region $\lvert m_S - m_h/2 \rvert \approx 0.1$ GeV. Accordingly, in contrast to its relevance for LHC constraints, the effect can be neglected in the computation of the relic density since considerably smaller couplings are required in the resonant regime. This is true even for the scenarios in which $S$ constitutes a subdominant fraction of DM considered in Sec.~\ref{sec:combined}.  
We neglect thermal corrections to the Higgs width and virtual corrections to the annihilation into an on-shell Higgs which have recently been computed in Ref.~\cite{Laine:2022ner}. However, these computations assume kinetic equilibrium. 

Finally, we mention an alternative mechanism to generate the DM density within the model. For very small (or \emph{feeble}) Higgs-portal couplings, DM may never thermalize with the SM bath. In this case, DM can be produced via freeze-in    ~\cite{McDonald:2001vt,Asaka:2005cn,Hall:2009bx}. Explaining the measured relic density in this scenario requires $\lambda_{HS}\sim 10^{-12}\!-\!10^{-11}$ for $m_S = (50-70)$ GeV. Due to the non-efficient annihilation processes in this regime, an initial abundance from processes taking place during reheating cannot be diluted as it is the case for thermalized DM discussed above. Hence, this scenario faces further constraints from physics of the very early Universe. Furthermore, the annihilation rate is way below the one providing a detectable signal in astroparticle data, such as the GCE (see Sec.~\ref{sec:combined}). We do not consider this case here.

\section{Collider searches}
\label{sec:collider}
Being independent of astrophysical uncertainties and cosmological assumptions, collider searches for invisible decays of the SM Higgs boson into $SS$ constitute a robust probe of the model. For $m_S<m_h/2$, the singlet scalar $S$ can contribute to the invisible width $\Gamma_{h,\rm{inv}}$ of the Higgs boson and we can directly reinterpret the limits derived by the experimental collaborations.
The invisible branching ratio is:
\begin{equation}
\mathcal{B}_{h,\rm{inv}}= \frac{\Gamma_{h,\rm{inv}}}{\Gamma_{h,\rm{inv}}+\Gamma_{h,\rm{SM}}}.
\end{equation}
The latest $95\%$ CL upper limits on $\mathcal{B}_{h,\rm{inv}}$ are 0.13~\cite{ATLAS-CONF-2020-008} (ATLAS, preliminary) and 0.19~\cite{CMS:2018yfx} (CMS, published). 

In this work, we are interested in the resonant region, $m_S\approx m_h/2$, where contributions of an off-shell Higgs boson production are relevant. Furthermore, as pointed out in Ref.~\cite{Heisig:2019vcj}, the total Higgs boson width can become a rapidly varying function of the invariant mass due to the opening of the Higgs boson decay channel into $SS$ requiring a computation beyond the fixed-width description.
This calls out for a more careful reinterpretation of the experimental results. To this end, we follow the method presented in Ref.~\cite{Heisig:2019vcj}.

Assuming the factorization of the Higgs boson production and decay channels, in an analogue way to Eq.~\eqref{eq:annisigma}, we can express the cross section for DM pair production at the LHC as
\begin{equation}
\label{eq:LHCDMprod}
\sigma_{pp\to SS} = \int \frac{\mathrm{d} s}{\pi} \,  \sigma_{h}(s) \, 
 |D_h(s)|^2 \, \sqrt{s}\, \Gamma_{h,\text{inv}}(s) \,\Theta(s - 4m_S^2)\,,
\end{equation}
where $\sigma_h(s)$ is the Higgs boson production cross-section for an (off-shell) Higgs boson with invariant mass $\sqrt{s}$.
This equals the on-shell production cross-section of a SM-Higgs-like scalar $\phi$ with mass $m_\phi$, $\sigma_h(s=m_\phi^2) = \sigma_\phi(m_\phi)$. 
\begin{figure}[t]
\includegraphics[width=\linewidth]{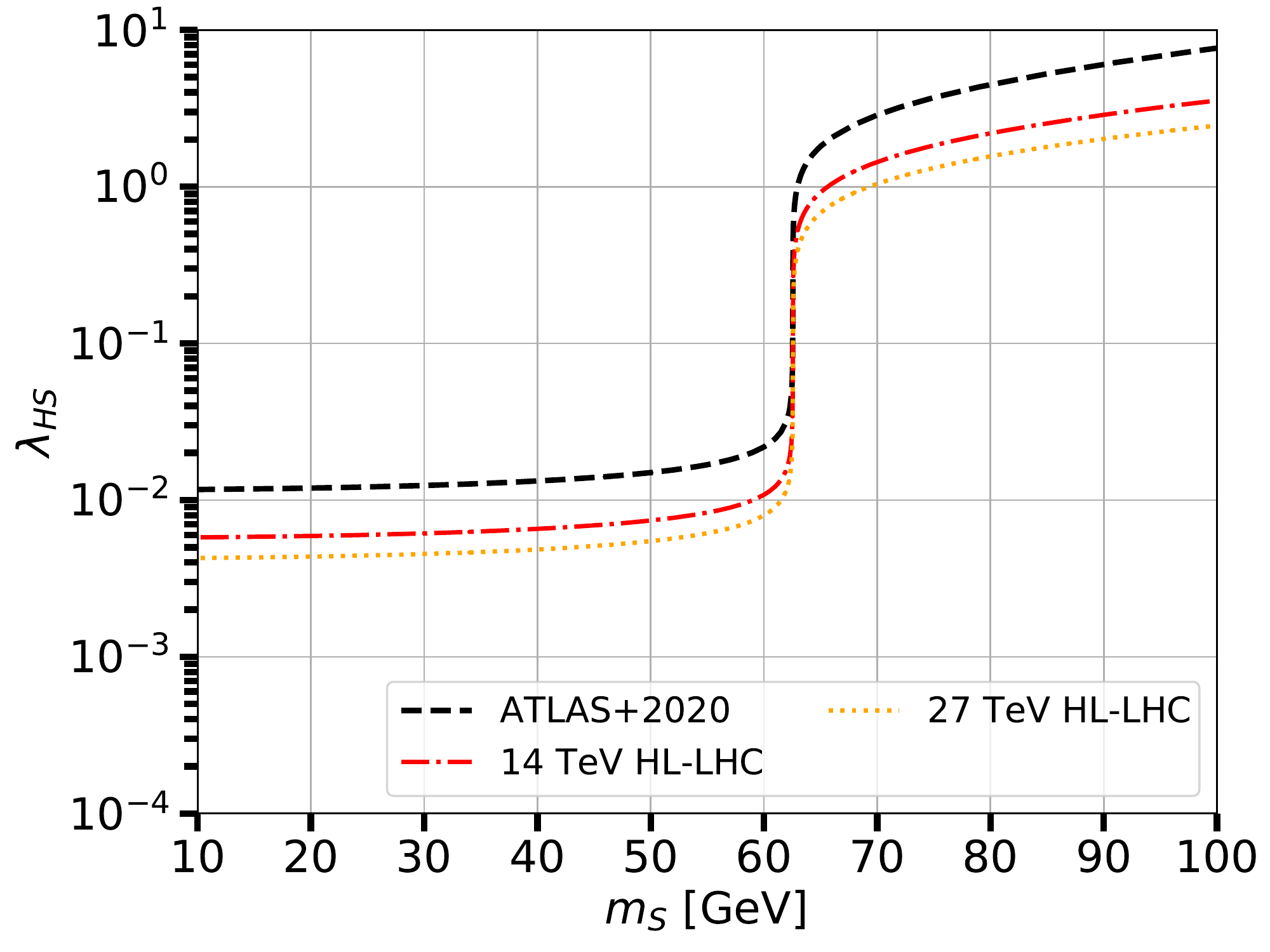}
\caption{$95\%$ CL upper limits for the parameter $\lambda_{HS}$ as a function of the DM mass $m_S$. We show the constraints obtained from a reinterpretation of the ATLAS search in Ref.~\cite{ATLAS-CONF-2020-008} and the projected limits for the HL-LHC and HE-LHC at 14 and 27 TeV, respectively, taken from~\cite{Heisig:2019vcj}.}
\label{fig:collider}
\end{figure}
Ref.~\cite{ATLAS-CONF-2020-008} reports a 95\% CL upper limit on the product of the cross section times the invisible branching ratio for such a scalar.
Using this result, we can employ Eq.~\eqref{eq:LHCDMprod} to obtain limits on the DM coupling in our model without the use of Monte Carlo simulations. We use the running width in the Higgs propagator, i.e.~replacing $m_h\to \sqrt{s}$ in Eq.~\eqref{eq:Gh} (see the appendix of Ref.~\cite{Heisig:2019vcj} for further details). 
We take the theoretical predictions for the cross section $\sigma_\phi(m_\phi)$ from Refs.~\cite{LHCHiggsCrossSectionWorkingGroup:2011wcg,LHCHXSWGBSM4:2016}. 

The result for the 95\% confidence level (CL) upper limit on $\lambda_{HS}$ is shown in Fig.~\ref{fig:collider} with a black, dashed curve, as a function of $m_S$.
For $m_S$ well below $m_h/2$ this result resembles the quoted upper limit on $\mathcal{B}_{h,\rm{inv}}$ of $0.13$, obtained in \cite{ATLAS-CONF-2020-008}, while it deviates strongly from the so-obtained limit for $m_S \gtrsim m_h/2$.
Notice that the constraint is about a factor of $2.5$ stronger than the one derived in Ref.~\cite{Heisig:2019vcj} on the basis of Ref.~\cite{CMS:2018yfx}.
We also show the potential for future improvement of the limit at the HL-LHC with 14 TeV and a HE-LHC upgrade with 27 TeV center-of-mass energy taken from Ref.~\cite{Heisig:2019vcj}. 

\section{Direct detection searches}
\label{sec:DD}
Direct detection experiments such as LZ and XENONnT can provide very tight constraints on the cross sections for the interactions of DM particle with nucleons \cite{XENON:2020kmp}.
Direct detection experiments typically measure upper limits on the elastic scattering cross-section off nucleons, as a function of the DM mass.
For the SHP model, the cross section is spin-independent only.
Both the XENONnT and the LZ experiments have recently released the tightest constraints so far on the spin-independent cross-section~\cite{XENON:2023sxq,LZ:2022ufs} given as a $90\%$ CL upper limit on $\sigma_{\mathrm{SI}}$ which are at the level of $10^{-47}$ cm$^2$ for DM masses of the order of few tens of GeV.
The upper limits have been derived assuming the Standard Halo Model.
The local DM density $\rho_{\odot}$ is conventionally fixed at $0.3$ GeV/cm$^3$, while DM velocity follows a Maxwellian velocity distribution, with velocity dispersion $220/\sqrt{2}$ km/s, escape velocity $v_{\rm{esc}} = 544$ km/s, and Earth velocity of $v_E = 232$ km/s. 
XENONnT and LZ are expected to reach upper limits at the level of $10^{-48}$ cm$^2$ \cite{XENON:2020kmp}.

The spin-independent cross-section in the framework of the SHP model is (see \cite{Cline:2013gha}):
\begin{equation}
\sigma_{\mathrm{SI}}=\frac{\lambda_{HS}^2}{4 \pi m_h^4} \frac{m_N^4 f_N^2}{\left(m_S+m_N\right)^2},
\end{equation}
where $m_N$ is the nucleon mass.
The term $f_N$ parametrizes the Higgs-nucleon interaction:
\begin{equation}
f_N=\sum_q f_q=\sum_q \frac{m_q}{m_N}\langle N|\bar{q} q| N\rangle,
\end{equation}
where the sum is made over all quark flavours $q$ with mass $m_q$.
A typical value of $f_N \approx 0.3$ is often adopted \cite{Cline:2013gha}.

To compute the spin-independent cross-section for the SHP model, we used two numerical codes: \maddm \cite{Backovic:2015cra} and \micromegas.
The results found with \micromegas are larger than the one of \maddm by a factor of $5\%$, independent from the DM mass.
The main features responsible for the difference are associated to the QCD corrections and to the contribution of higher twist operators, which are not taken into account in \maddm\footnote{We verified that turning off the running of the QCD coupling in \micromegas (setting the variable {\tt qcdNLO=1} in the script {\tt directDet.c}), and turning off the contributions from higher twist operators (setting {\tt Twist2On=0}), the two codes give results which are compatible within $1\%$.}.

\begin{figure*}[t]
\includegraphics[width=0.49\linewidth]{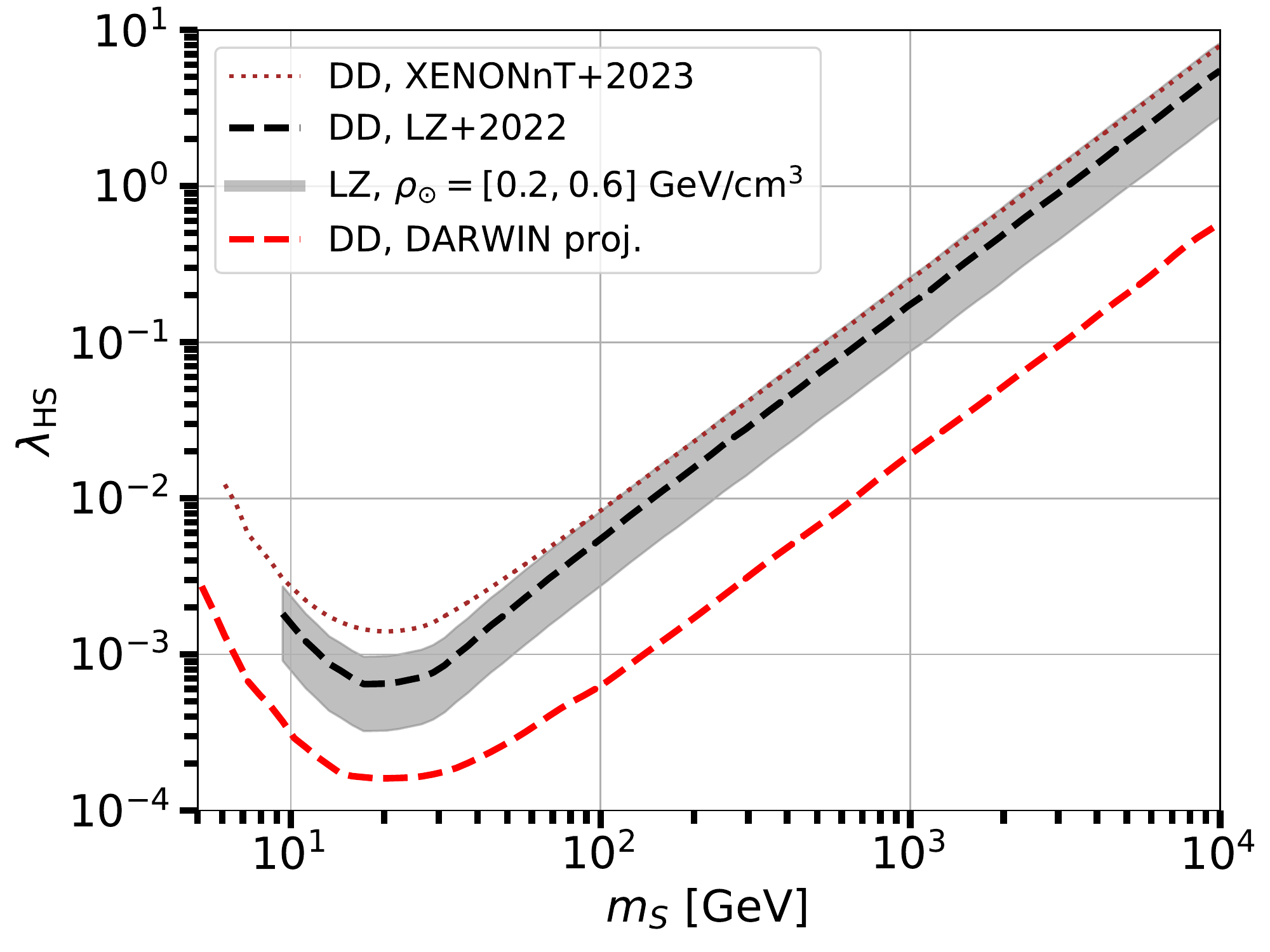}
\includegraphics[width=0.49\linewidth]{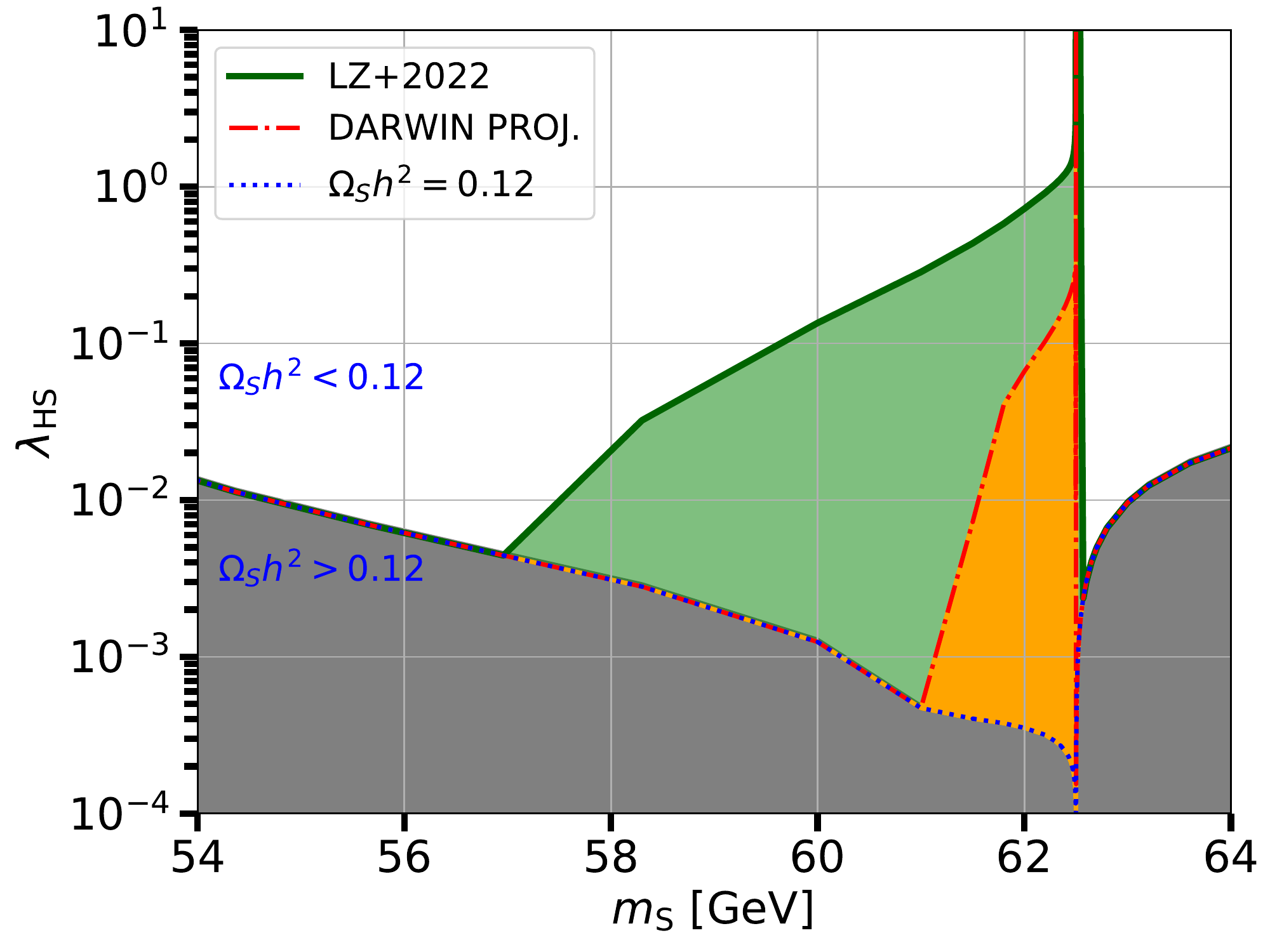}
\caption{Left panel: The $90\%$ CL upper limits for the parameter $\lambda_{HS}$ as a function of the DM mass $m_S$ obtained with direct detection experiments. We recast the exclusion limits of LZ~\cite{LZ:2022ufs} and XENON1T~\cite{XENON:2023sxq} along with the projected sensitivity for the future experiments DARWIN \cite{DARWIN:2016hyl} for the SHP model. The grey region represents the uncertainty in the upper limit induced by the uncertainty on the local DM density $\rho_{\odot}$, which we vary in the range (0.2\,--\,0.6) GeV/cm$^3$.
Right panel: Same as left but with the constraints re-scaled on the basis of the actual value of $\Omega_S h^2$ following the assumption that the DM might be subdominant or overabundant, i.e.,~by using Eq.~\eqref{eq:direct_detection_upper_limit_fraction_relic_density}.
We show the allowed region obtained using the upper limits from the LZ experiment \cite{LZ:2022ufs} (green region) and the projected sensitivity for the future DARWIN experiment \cite{DARWIN:2016hyl} (orange region). We also show the combination of $\lambda_{HS}$ and $m_S$ that provides the correct relic abundance and we display the region where the particle $S$ is overabundant with a grey region.}
\label{fig:DD}
\end{figure*}

In Fig.~\ref{fig:DD} we show the upper limits found using the available data from the LZ \cite{LZ:2022ufs} and XENONnT \cite{XENON:2023sxq} experiments.
In particular, LZ can probe values of the coupling parameter $\lambda_{HS}$ as low as about $10^{-3}$ for DM masses of a few tens of GeV.
However, the local DM density profile and its velocity distribution is characterized by sizable uncertainties.
For DM masses below $10$ GeV, the uncertainties on the velocity distribution play an important role, while the ones on the value $\rho_{\odot}$ are the dominant source of uncertainty at higher DM masses  (see \cite{Wu:2019nhd}).
Since the parameter space we are considering consists of DM masses above 10 GeV, we take into account only the uncertainties on $\rho_{\odot}$, and we consider it to vary in the range (0.2\,--\,0.6) GeV/cm$^3$~\cite{Salucci:2010qr,Nesti:2013uwa,Benito:2019ngh,2020MNRAS.495.4828G}.
In Fig.~\ref{fig:DD}, left panel, we show the uncertainty band associated to $\rho_{\odot}$.
In the same plot we also show the projected sensitivity for the future experiment DARWIN \cite{DARWIN:2016hyl}, which will reduce the upper limits by nearly a factor of 4 for the considered mass range.

Assuming that the total DM relic density can be associated to only one single DM candidate, in our case the $S$ scalar, the differential rate of detection $dR/dE$ is proportional to $\sigma_{S N}^{\mathrm{SI}} \rho_{\odot}/m_S$, where $\rho_{\odot}$ is the local DM mass density. 
On the contrary, assuming that the candidate $S$ constitutes only a fraction of the full DM component, i.e.~that the relic density of the particle $S$ is a fraction of the total DM relic density, we have to apply an appropriate rescaling to the cross section upper limit.
Indeed, in this case, the local DM density of the DM candidate $S$ is lower, and can be simply assumed to scale as:
\begin{equation}
    \xi = \frac{\Omega_S h^2}{\Omega_{\rm{DM}}h^2}
    \label{eq:relic_density_fraction}\,.
\end{equation}
Assuming that there is no difference in how the particle $S$ and the other DM components cluster in the Galaxy, the local energy density of $S$ can be written as \begin{equation}
    \rho_\odot(S) = \xi \cdot \rho_{\odot}.
\end{equation}
As a consequence, the upper limits on $\sigma_{\mathrm{SI}}$ associated to $S$ can be computed as the parameter space of $m_S,\lambda_{HS}$, satisfying the condition:
\begin{equation}
\xi \cdot \sigma_{\mathrm{SI}}(m_S,\lambda_{HS}) \leq \sigma^{\rm{UL}}_{\rm{EXP}},
\label{eq:direct_detection_upper_limit_fraction_relic_density}
\end{equation}
where $\sigma^{\rm{UL}}_{\rm{EXP}}$ is the $90\%$ CL upper limit of the experiment under consideration.
In Fig.~\ref{fig:DD}, right panel, we show the results obtained by including the effects of considering a fraction of DM into the calculation of the relic density.
Using the LZ upper limit, the only region allowed corresponds to $S$ masses in the range $(57\!-\!62.51)$ GeV.
Additionally, we also consider the projected DARWIN upper limits and the remaining narrow region is $m_S=(61 \, \mathrm{GeV}-m_h/2)$, i.e~below $m_h/2$.
Over the resonance region, the couplings compatible with the direct detection constraints are much larger than what is shown in Fig.~\ref{fig:DD}.
For example, for $m_S=60$ GeV, the upper limits on the coupling parameter increases by a factor of about 100.
Therefore, the requirement of achieving the correct relic density for the $S$ particles in the calculation of the direct detection has the consequence of restricting the mass range compatible with the data upper limits but it provides weaker constraints on the $\lambda_{HS}$ parameter in the same mass range.
Future experiments seem to not be able to rule out values of $\lambda_{HS}$ smaller than 0.1 if the DM mass remains slightly below $m_h/2$.

\section{Indirect detection}
\label{sec:ID}
Current experiments measuring cosmic particles, such as $\gamma$ rays, neutrinos or CRs, count the indirect detection of a possible DM signal among their most important science cases (see, e.g.,~\cite{Fermi-LAT:2016afa} for {\it Fermi}-LAT).
In particular, these experiments focus their searches on the detection of fluxes of the rarest astrophysical particles such as: photons (from radio to $\gamma$ rays), neutrinos and antineutrinos $\nu$, antiprotons $\bar{p}$, positrons $e^+$ and antinuclei.
Among these particles, photons and, to an even larger extent, $\nu$ have the great advantage to travel almost undisturbed across the Universe and are able to provide directional information. This allows to focus the DM search in the direction of astrophysical sources which are expected to have very large density of DM.
Among the most promising targets, we mention the Milky Way's Galactic center, dSphs and the clusters of galaxies.
Current experiments can detect astrophysical neutrinos for energies above 1 TeV and with quite low statistics (see, e.g.,~\cite{IceCube:2021uhz}). These two aspects make them not suitable for our scopes. 
Concerning photons, we focus on the $\gamma$ rays detected by {\it Fermi}-LAT and we use the data available for the GCE from Refs.~\cite{DiMauro:2021raz,Cholis:2021rpp} and the recent analysis of DM searches in dSphs from Refs.~\cite{DiMauro:2021qcf,DiMauro:2022hue}.
Finally, we consider among the CRs only antiprotons because the SHP predicts a large hadronic production of these particles, while the $e^{\pm}$ production is less important. Finally, we won't consider cosmic antinuclei because currently no firm detection has been published so far and upper limits are much weaker than other probes (see, e.g.,~\cite{Korsmeier:2017xzj}). 

\subsection{The \texorpdfstring{$\gamma$}{\textgamma} rays from the Galactic center and dwarf galaxies}
\label{sec:gamma}

\subsubsection{Model}

The $\gamma$-ray emission from DM particle interactions includes two components: the direct and indirect production.
The first one is also called prompt emission and it is due to the direct production of $\gamma$ rays through an intermediate annihilation channel. 
The prompt emission is theoretically calculated as follows:
\begin{equation}
\frac{dN}{dE d\Omega} = \frac{1}{2} \frac{r_{\odot}}{4\pi}  \left( \frac{\rho_{\odot}}{m_{S}} \right)^2 \bar{\mathcal{J}} \times  \langle \sigma \vMol \rangle \sum_f \frac{\langle \sigma v \rangle_f}{\langle \sigma v \rangle}   \left( \odv{N_\gamma}{E} \right)_f,
\label{eq:fluxprompt}
\end{equation}
where $\rho_{\odot}$ is the local DM density, and $r_{\odot}$ is the distance of the Earth from the center of the Galaxy.
We use $r_{\odot} =8.12$ kpc, as measured recently in Ref.~\cite{2019ApJ...871..120E}. 
The term $\bar{\mathcal{J}}$ is the geometrical factor averaged over the solid angle $\Delta \Omega$ spanned by the region of interest (ROI) considered in the analysis.
This quantity is calculated as the integral performed along the line of sight (l.o.s.) $l$ of the squared DM density distribution $\rho^2$ normalized by $\Delta \Omega$:
\begin{equation}
\bar{\mathcal{J}} = \frac{1}{\Delta \Omega}  \int_{\Delta \Omega} d\Omega \int_{l.o.s.}  \frac{dl}{r_{\odot}}  \frac{\rho^2[r(l,\Omega)]}{\rho^2_{\odot}} .
\label{eq:geom}
\end{equation}

\begin{figure*}[th]
\includegraphics[width=0.49\linewidth]{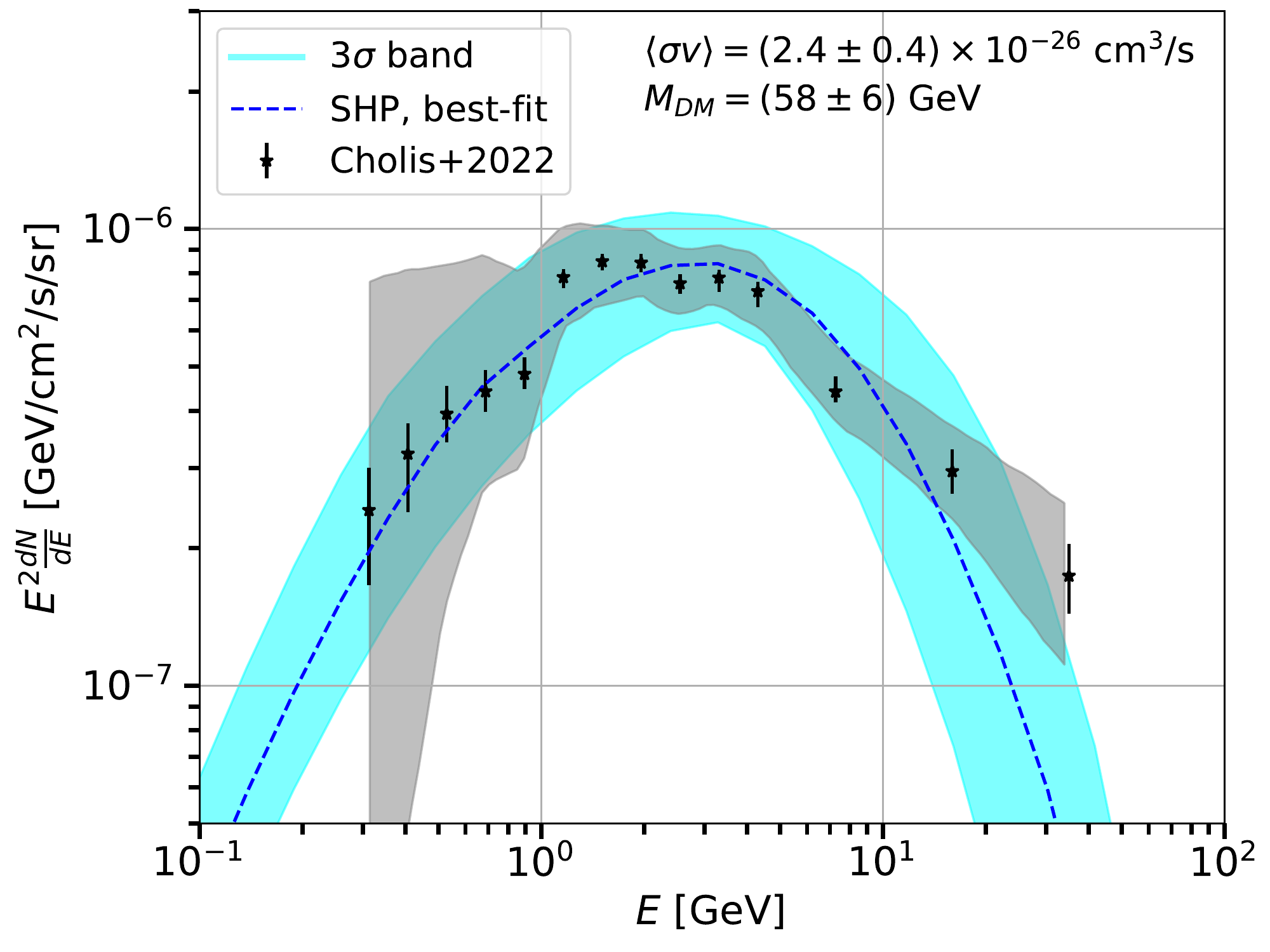}
\includegraphics[width=0.49\linewidth]{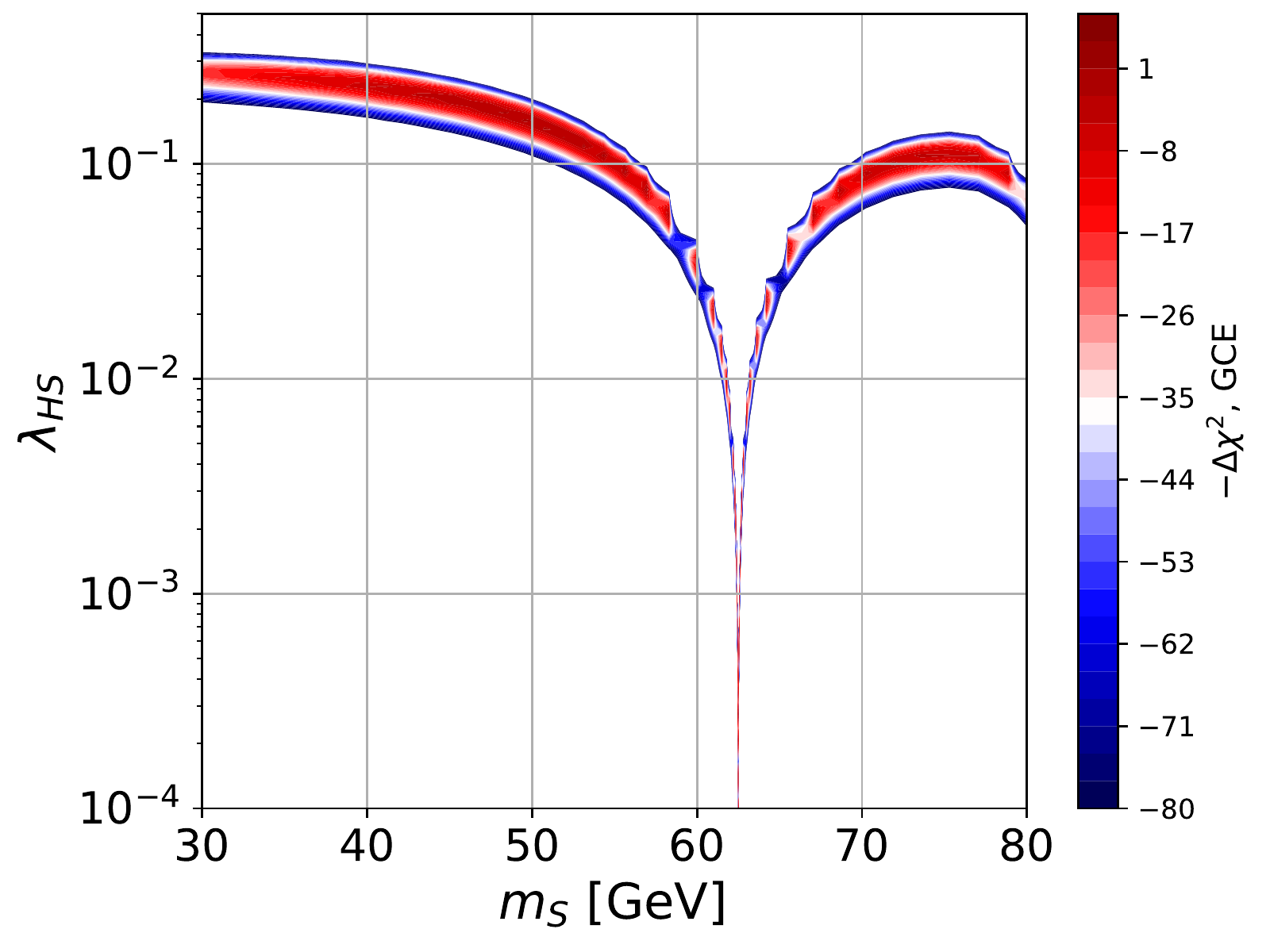}
\caption{Left panel: Best-fit SED for DM in the SHP model to the GCE found in Ref.~\cite{Cholis:2021rpp}. We show the data with the statistical and systematic errors (grey band) and the theoretical best-fit (blue curve) with the $3\sigma$ uncertainty band. Right panel: Contour plot for the $-\Delta \chi^2$ defined as $-(\chi^2(m_S,\lambda_{HS})-\chi^2_{\rm{min}})$, where $\chi^2_{\rm{min}}$ is the chi-square of the best fit.}
\label{fig:GCE}
\end{figure*}

The term $(dN_\gamma/d E)_f$ is the $\gamma$-ray spectrum from DM annihilation for a specific annihilation channel labeled as $f$ and $\langle \sigma v \rangle_f/\langle \sigma v \rangle$ is its relative contribution of the annihilation cross section into a specific channel $f$ with respect to the total annihilation cross section. 
As explained in Sec.~\ref{sec:spectra} we calculate the source spectra in the SHP model using \maddm.

In case of DM particles annihilating into leptonic channels, i.e. $e^+ e^-$, $\mu^+ \mu^-$ and $\tau^+ \tau^-$, there is a secondary production of $\gamma$ rays that could be relevant.
This involves $e^{\pm}$ produced from the prompt emission that can subsequently generate $\gamma$ rays through inverse Compton scattering on the interstellar radiation fields photons.
This component is particularly relevant for the Galactic center  where the density of the starlight and dust components of the interstellar radiation fields are roughly a factor of 10 higher than their local density (see, e.g.,~\cite{Porter_2008}).
As demonstrated in Ref.~\cite{DiMauro:2021qcf}, hadronic annihilation channels have a contribution from inverse Compton scattering relevant only for energies below 1 GeV, for $m_S$ between tens of GeV up to 100 GeV, which is the most interesting region in our analysis. For these energies, the uncertainties in the GCE flux are very large and the significance of detection is small. Therefore, we decide to neglect the contribution of secondary $\gamma$ rays coming from DM annihilation. 

The exact DM density distribution in the center of the Galaxy is not well known. In Ref.~\cite{DiMauro:2021qcf}, the authors considered an hybrid approach to estimate the uncertainty on the density profile.
They took into account two factors: the data on Galaxy rotation curve, which constrains $\rho$ beyond a few kpc from the Galactic center; and the GCE data, which fixes the density shape between about 0.1 and few kpc under the assumption that the GCE is originated by DM annihilation. 
Using this strategy they identified three models, labeled as MIN, MED MAX, that bracket the uncertainty on the geometrical factor. In particular, they have found that the variation for the $\bar{\mathcal{J}}$ value between the MIN and MAX models is about a factor of 7.
We considered this uncertainty in our results.
Instead, the uncertainty on $\bar{\mathcal{J}}$ for dSphs is included directly in the analysis of the $\gamma$-ray data from these objects. In fact, this parameter is treated as a nuisance parameter of the likelihood fit, using a gaussian prior with an average taken as the observed geometrical factor and with the $1\sigma$ error as the width (see \cite{DiMauro:2021qcf,DiMauro:2022hue} for more details).

\subsubsection{Results}
\label{sec:resgamma}

We first provide the results of the fit to the GCE data found in \cite{DiMauro:2021raz,Cholis:2021rpp}.
We perform a fit to the data taking into account statistical and systematic errors.
The latter include the uncertainties due to the choice of the Galactic interstellar emission model.
The DM mass $m_S$ and the annihilation cross section $\langle \sigma \vMol\rangle$ are free parameters of the fit.
The left panel of Fig.~\ref{fig:GCE} shows the result of the fit to the data found in \cite{Cholis:2021rpp}, which gives the best-fit values for the DM parameters: $m_S = (58\pm6)$ GeV and $\langle \sigma \vMol\rangle = (2.4\pm0.4) \times 10^{-26}$ cm$^3$/s. The best-fit value for the coupling parameter is between 0.03--0.18. It varies significantly because the best-fit masses are close to the Higgs resonance for which the $\langle \sigma \vMol \rangle$ increases dramatically, and the required coupling decreases.
This is visible in the right panel of Fig.~\ref{fig:GCE}, where we show the $\chi^2$ profile as a function of the parameters $m_S$ and $\lambda_{HS}$.
When $m_S \approx 50$ GeV, the value of $\lambda_{HS}$ that fits the data is approximately $(1\!-\!2) \times 10^{-2}$. 
Close to the resonance $m_h/2$, the best-fit of the coupling decreases to values of the order $(1\!-\!2) \times 10^{-4}$. This is due to the fact that by fixing $\lambda_{HS}$, the $\langle \sigma \vMol\rangle$ increases significantly when $m_S \approx m_h/2$.
For masses around $(70\!-\!80)$ GeV, the best-fit value of the coupling increases to values of the order $10^{-1}$.
The $\chi^2$ profile we find using the data released in the other reference considered in this paper for the GCE, i.e.~Ref.~\cite{DiMauro:2021raz}, is similar to the one shown in Fig.~\ref{fig:GCE} and obtained from the GCE SED obtained in Ref.~\cite{Cholis:2021rpp}.
As a final remark, we remind that the geometrical factor for the Galactic center is not well known. As seen before, in Ref.~\cite{DiMauro:2021qcf} an uncertainty of about a factor of $7$ in the $\bar{\mathcal{J}}$ value has been estimated between the models describing the GCE density profile. This translates into a systematic also in the best-fit of $\lambda_{HS}$.
In particular, the calculation of the $\gamma$-ray flux is proportional to $\langle \sigma \vMol \rangle \cdot \bar{\mathcal{J}}$.
That means a variation of $\bar{\mathcal{J}}$ can be reabsorbed by an opposite variation of the annihilation cross section, namely in a variation of the value of the coupling $\lambda_{HS}$, where we have $\langle \sigma \vMol \rangle \propto \lambda^2_{HS}$.
Therefore, a systematic of a factor of 7 in the geometrical factor causes a systematic in the value of $\lambda_{HS}$ of about $\sqrt{7}\approx 2.6$.

\begin{figure*}[t]
\includegraphics[width=0.49\linewidth]{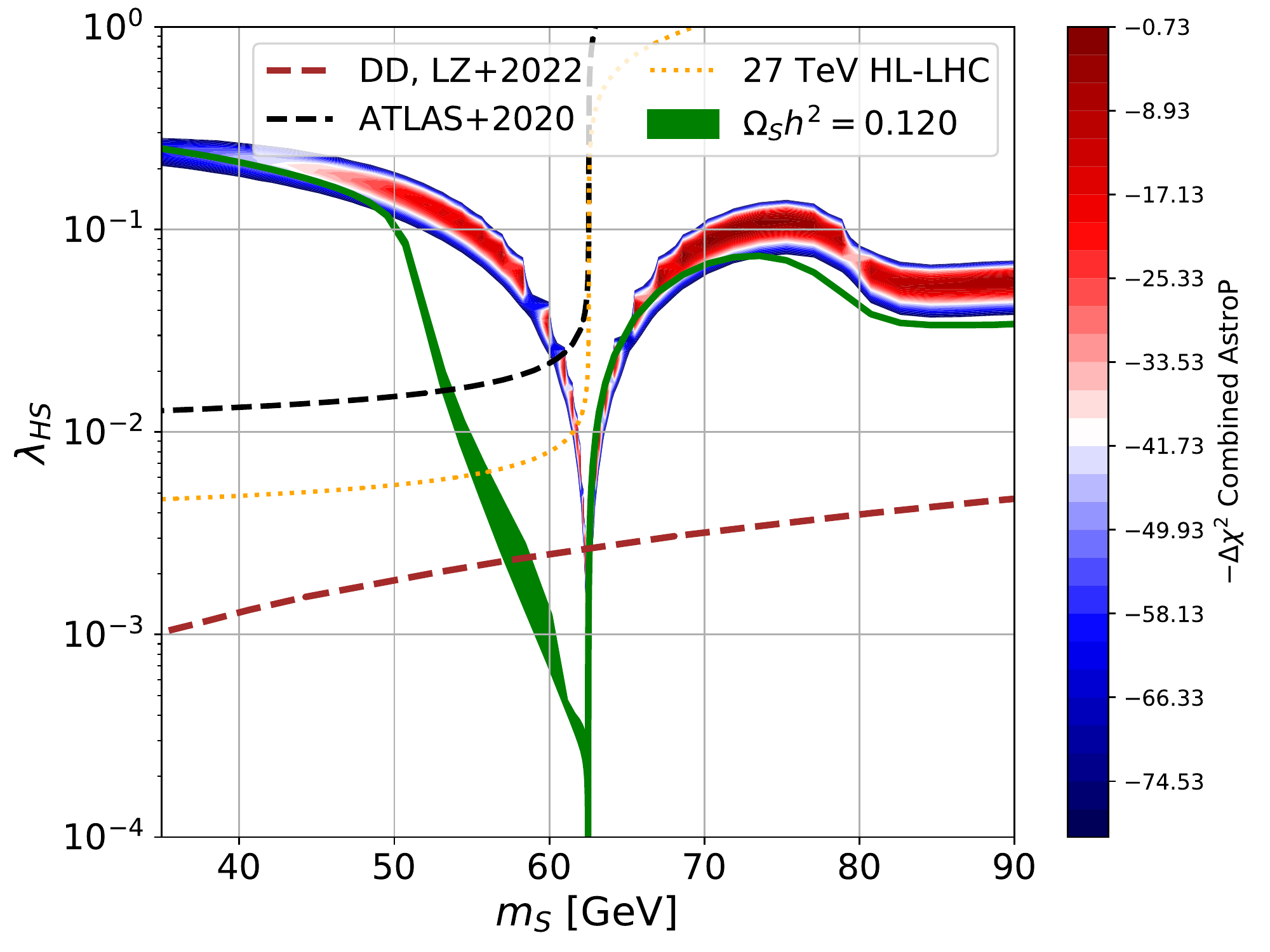}
\includegraphics[width=0.49\linewidth]{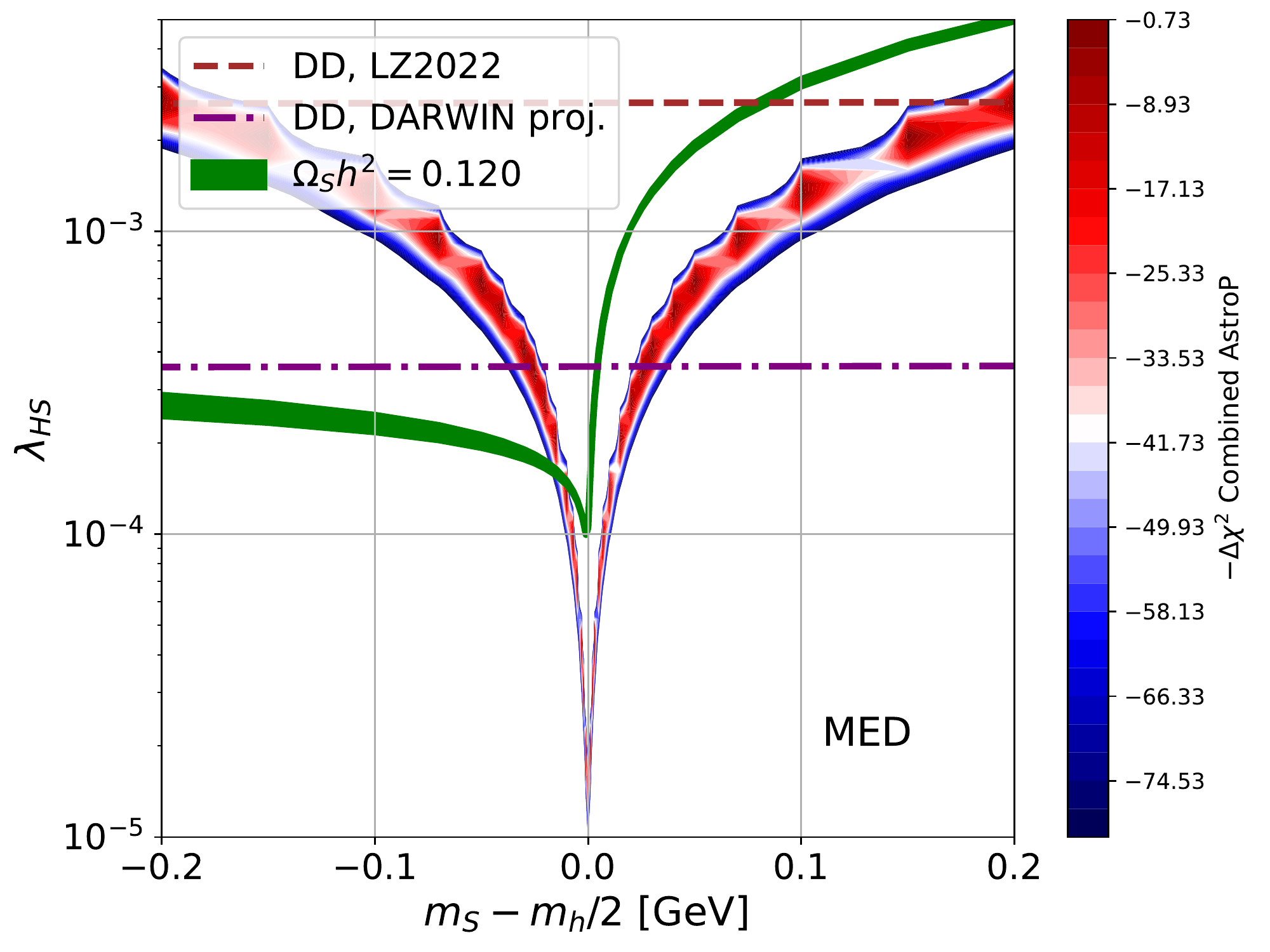}
\caption{Left panel: Contour regions obtained with a combined fit to cosmic particle flux data (GCE, dSphs and antiprotons). We also show the upper limits on $\lambda_{HS}$ obtained from ATLAS data \cite{ATLAS-CONF-2020-008} (red dashed line) and the HL-LHC projections for 27 TeV \cite{Heisig:2019vcj} (orange dotted line). Direct detection upper limits refer to LZ data \cite{LZ:2022ufs} (brown dashed line) and projections to DARWIN (purple dot-dashed). We also report the region of the parameter space compatible with the observed DM relic abundance via thermal freeze-out (green region) in case of model {\tt fBE}, including the uncertainty coming from the different choice of the QCD correction approach, labelled with {\tt QCD A} and {\tt QCD B}. Right panel: Same as in the left panel but for the region around the resonance $m_S \approx m_h/2$.} 
    \label{fig:combinedallMED}
\end{figure*}

\begin{figure*}[t]
\vspace{3.5ex}
\includegraphics[width=0.49\linewidth]{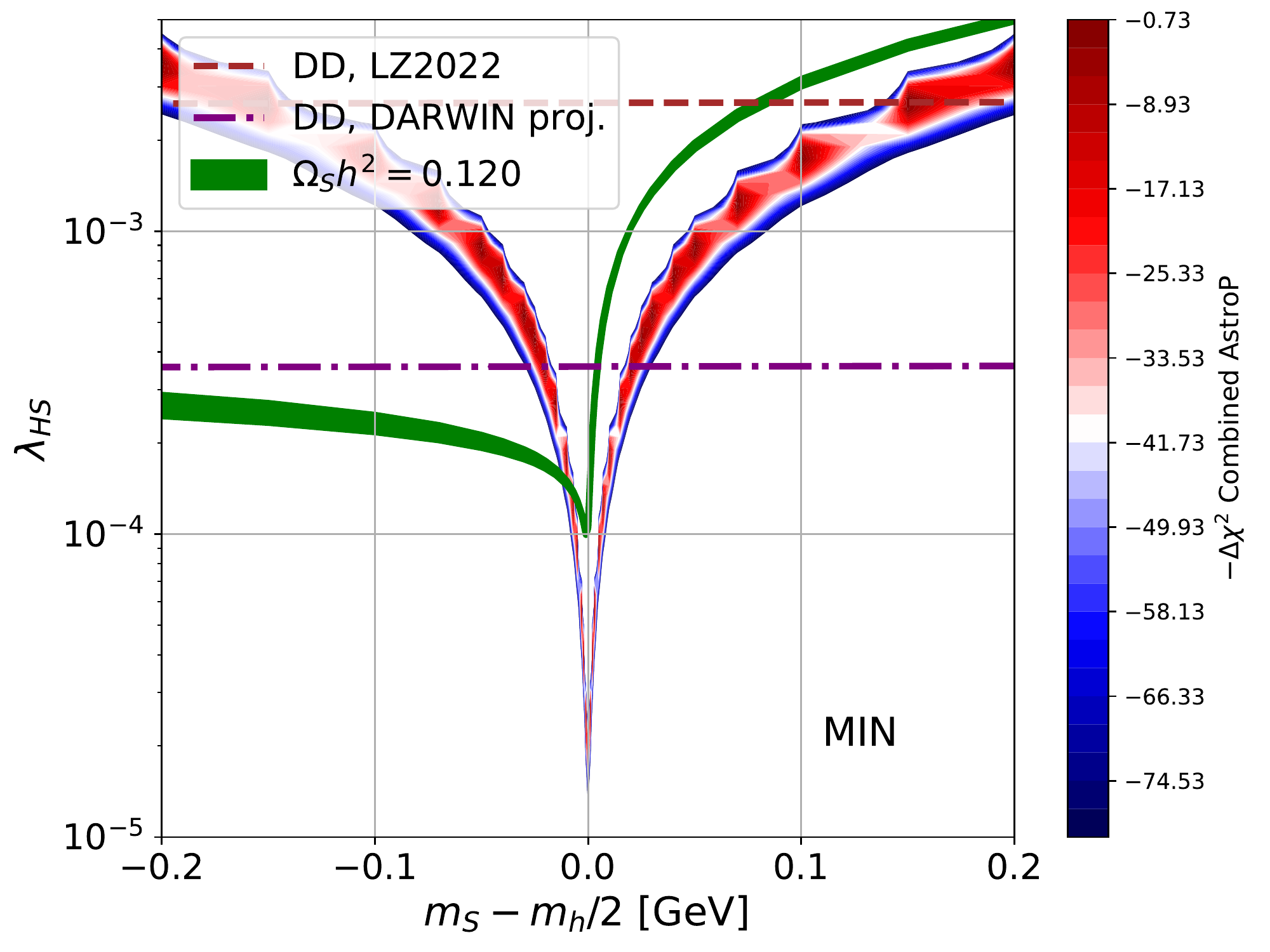}
\includegraphics[width=0.49\linewidth]{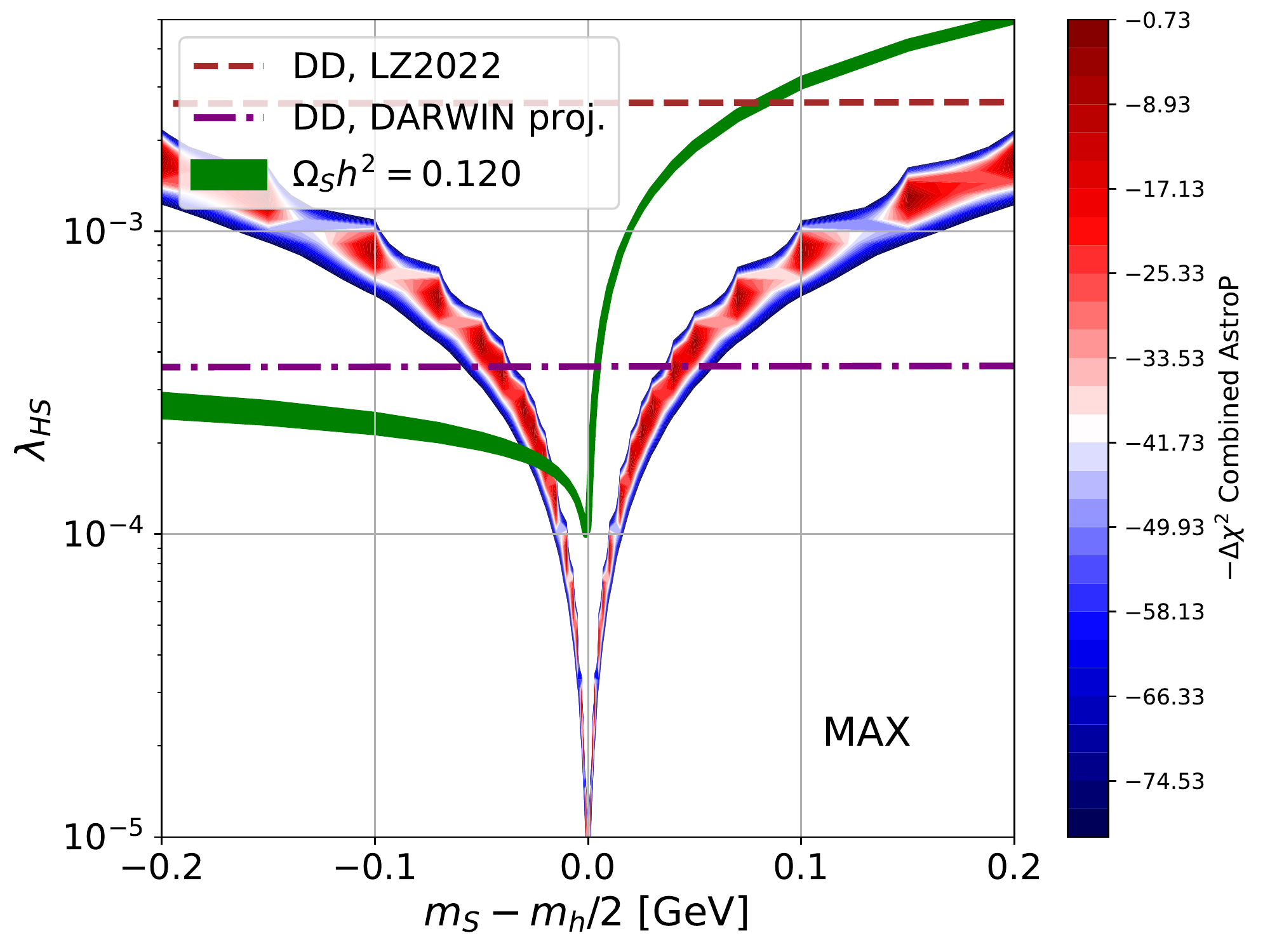}
\caption{Same as in the right panel of Fig.~\ref{fig:combinedallMED} but considering the MIN (left panel) and MAX (right panel) DM density parameters for the GCE analysis.}
\label{fig:combinedallMINMAX}
\end{figure*}

We also analyzed the $\gamma$-ray constraints coming from dSphs. In particular we use the results of Refs.~\cite{DiMauro:2021qcf,DiMauro:2022hue}. In the former, a sample of 48 dwarf galaxies, both classical and ultrafaint, have been analyzed using 11 years of data and assuming that the objects are pointlike (i.e. smaller than the Fermi-LAT point-spread function). In the latter, 12 years of data in the direction of 22 dSphs have been analyzed considering a spatial extended templates for the DM flux.
In both cases no clear detection has been found with a maximum value of the test statistic of the order of 9.
Using the data released in Refs.~\cite{DiMauro:2021qcf,DiMauro:2022hue} we determined the $\chi^2$ profile as a function of the $m_S$ and $\langle \sigma \vMol\rangle$ as we did for the GCE.
The peak of the signal is located at masses around $(50\!-\!300)$ GeV and $\langle \sigma \vMol \rangle \approx (0.4\!- \!6) \times 10^{-26}$ cm$^3$/s.

\subsection{Antiproton flux}
\label{sec:pbar}

DM annihilation can induce a primary flux of $\bar{p}$.
The source term which denotes the differential production rate of $\bar{p}$ per volume, time and energy reads as:
 \begin{equation}
\mathcal{Q}(E,\vec{r}) = \left( \frac{\rho(\vec{r})}{\rho_{\odot}} \right)^2 \sum_i \frac{\langle \sigma v \rangle_i}{\langle \sigma v \rangle}   \left( \odv{N_{\bar{p}}}{E} \right)_i,
\label{eq:Q}
\end{equation}
where $(\odv{N_{\bar{p}}}/{E})_i$ is the spectrum of $\bar{p}$ produced from DM particle interactions and it depends on the specific annihilation channel assumed and labeled with $i$.
In addition, there is an astrophysical antimatter background which originates from the scattering of CR protons and nuclei on the interstellar matter.

To properly calculate the flux of secondary and DM antiprotons one should calibrate the galactic propagation parameters such that the spectra of primary particles reproduce AMS-02 observations.
We use the results presented in Ref.~\cite{Kahlhoefer:2021sha} where the authors have first fitted the data sets of proton, helium, and the antiproton-to-proton ratio from AMS-02 collected over 7 years \cite{AGUILAR2020} and complement these data sets with low-energy data for protons and helium from the Voyager satellite \cite{Stone150}.
Their propagation model includes diffusion (parametrized by a smoothly broken power-law in terms of the particle rigidity), reacceleration, energy losses, secondary production/fragmentation processes, and solar modulation modelled is a force field approximation (see Ref.~\cite{Kahlhoefer:2021sha} for details).
By adding a possible DM contribution to the antiproton source spectra, our analysis provides a $\chi^2$ profile as a function of the parameters $m_S$ and $\lambda_{HS}$ of the SHP model.
We combine these results for $\bar{p}$, published in Ref.~\cite{Kahlhoefer:2021sha}, with the ones found for $\gamma$ rays from the Galactic center and dSphs.
The authors recently followed up with an updated analysis \cite{Balan:2023lwg}.
In that work, they have tested two models, labelled as {\tt INJ.BRK} and {\tt DIFF.BRK}.
The former describes the injection spectra of the primary CRs with a broken power law using different slopes for $p$ and He.
The diffusion coefficient is modeled as a broken power law with a single break.
The latter model, instead, utilizes a single power law for the spectra, while the diffusion coefficient has both a low and an high-energy break.
In this analysis, the authors perform fits to the AMS-02 antiproton data with and without including the possible correlations~\cite{Heisig:2020nse} in the data.
The {\tt INJ.BRK} model provides evidence for a possible DM contribution similarly to what has been found in Refs.~\cite{Korsmeier:2017xzj,Korsmeier:2018gcy,Kahlhoefer:2021sha}.

For completeness, we considered both models in our analysis.
In particular, the {\tt INJ.BRK} model results in very minor changes on our constraints, while the {\tt DIFF.BRK} model shows that there is no evidence for a DM contribution for masses below 100 GeV.
As a consequence, the latter model allows us to impose tight constraints on the SHP model parameter space.
We will discuss this possibility in Sec.~\ref{sec:SHP_UL}.

\section{Combined results from direct, indirect, collider searches and cosmology}
\label{sec:combined}

\subsection{The \texorpdfstring{$S$}{S} particle as dominant dark matter component}
\label{sec:Sall}

In this section, we combine the results found previously using $\gamma$-ray and $\bar{p}$ cosmic data, collider, direct detection constraints and cosmological measurements.
We first combine the results obtained using $\gamma$-ray and $\bar{p}$ flux data by summing the $\chi^2$ profile defined as a function of the parameters $m_S$ and $\lambda_{HS}$.
In Fig.~\ref{fig:combinedallMED} we show the best-fit region from astroparticle data compared with the constraints from collider, direct detection and relic density.
For the direct detection we show the upper limits on $\lambda_{HS}$ obtained with the LZ data (see Sec.~\ref{sec:DD}) while for collider searches we use the upper limits found with CMS data for the vector boson fusion (see Sec.~\ref{sec:collider}).
In this figure as well as in Fig.~\ref{fig:combinedallMINMAX}, we assume that the $S$ particles make up $100\%$ of the DM relic density. This assumption enters the calculation of the indirect and direct detection signals whereas the green band shows the coupling that provides the respective relic density through thermal freeze-out assuming a standard cosmological history.
There are two intersections of the green and color-shaded bands indicating compatibility with the relic density and astroparticle data, respectively. The first one is at a DM mass $m_S\approx m_h/2$ and $\lambda_{HS} \approx (1.4\!-\!1.7) \times 10^{-4}$. This is compatible with the collider and direct detection constraints. The second one is for $m_S = (63-69)$ GeV and $\lambda_{HS} = (2-6)\times 10^{-2}$. However, this part of the parameter space is ruled out by direct detection constraints. 
Therefore, the only region of the parameter space that fits well cosmic data, compatible with relic density and consistent with collider and direct detection constraints is for $m_S \approx m_h/2$, to be precise $(10\!-\!20)$ MeV smaller than $m_h/2$, and $\lambda_{HS}= (1.4\!-\!1.7) \times 10^{-4}$.
In this region of the parameter space the GCE is fitted with a reduced $\chi^2$ of about 0.8, which implies that the fit is statistically appropriate.

As explained in Sec.~\ref{sec:gamma}, there is a quite large uncertainty related to the exact DM density in the center of the Milky Way, which in Ref.~\cite{DiMauro:2021qcf} has been estimated to be included among the MIN, MED and MAX models. In Fig.~\ref{fig:combinedallMINMAX}, we show the combined results obtained with the MIN and MAX models observing that a change of the DM density in the center of the Galaxy has little to no effect on our conclusions. With these models, there is still a region very close to the Higgs resonance which fits well cosmic fluxes data and which is compatible with the collider and direct detection experiments.
The only quantity that changes is the value of $\lambda_{HS}$ that fits the cosmic fluxes data and for which we have the correct relic abundance. By considering the MIN and MAX model the coupling can vary in the range $(1.2\!-\!2.0) \times 10^{-4}$.

\begin{figure*}[t]
\includegraphics[width=0.49\linewidth]{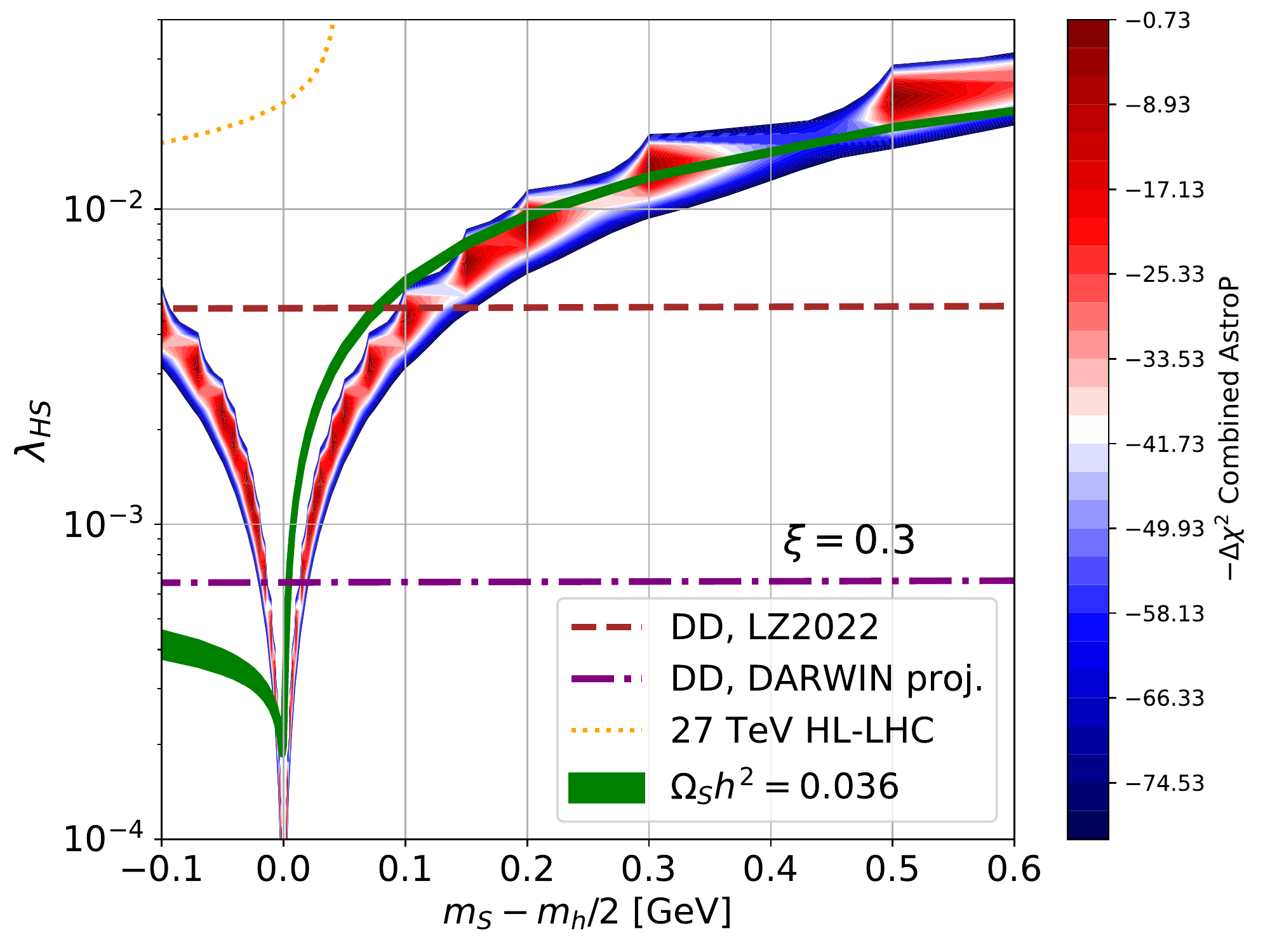}
\includegraphics[width=0.49\linewidth]{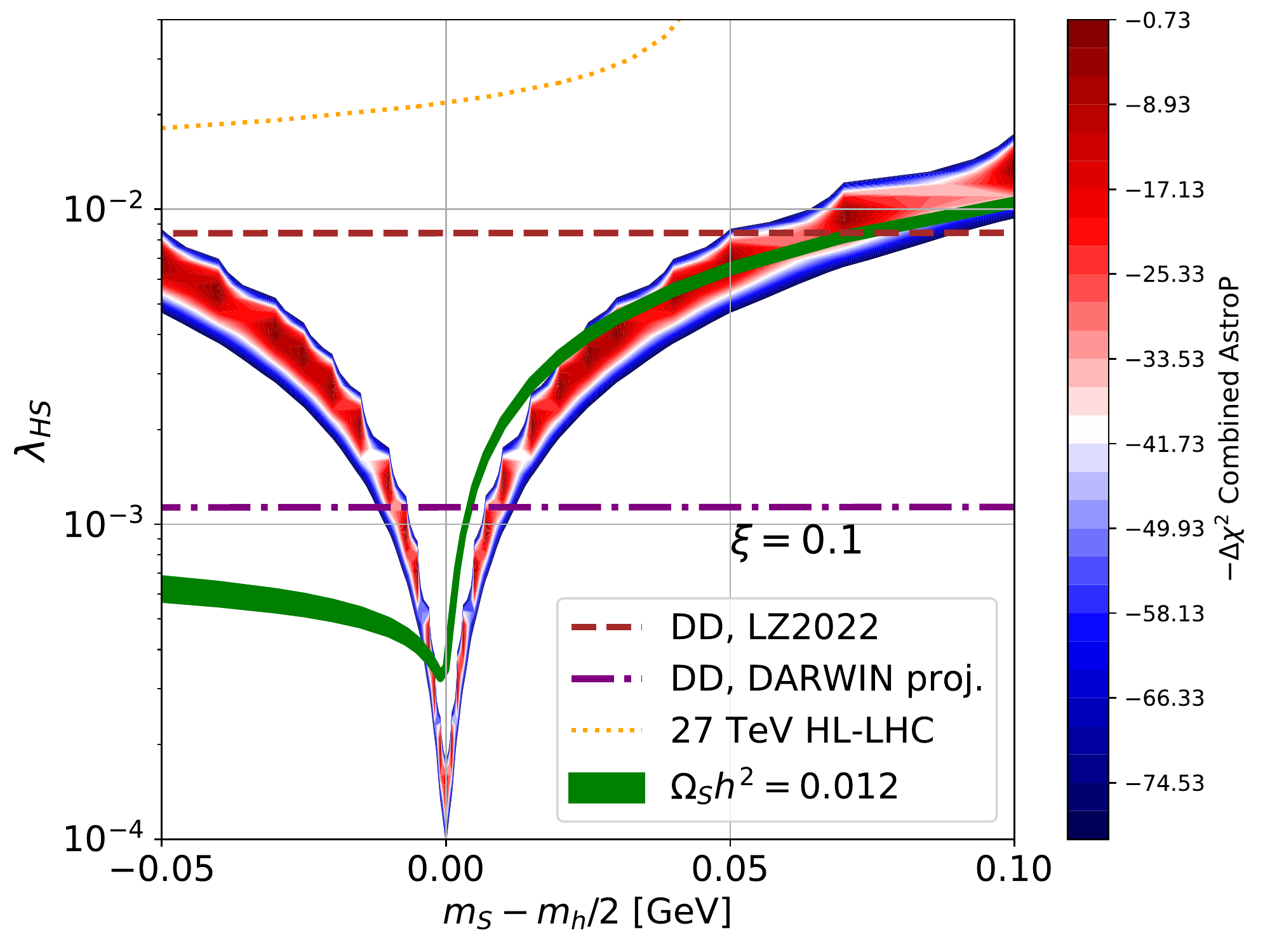}
\includegraphics[width=0.49\linewidth]{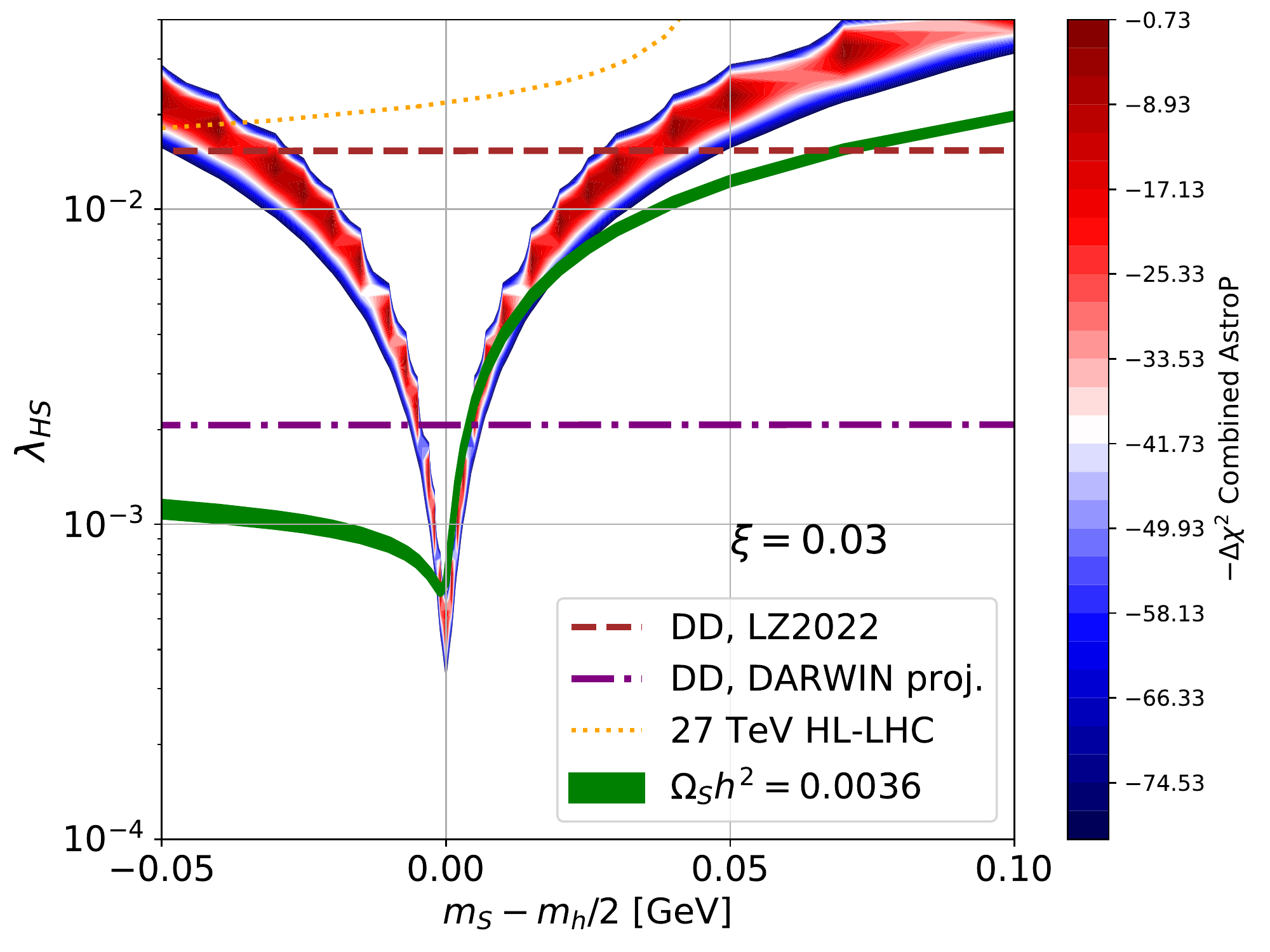}
\includegraphics[width=0.49\linewidth]{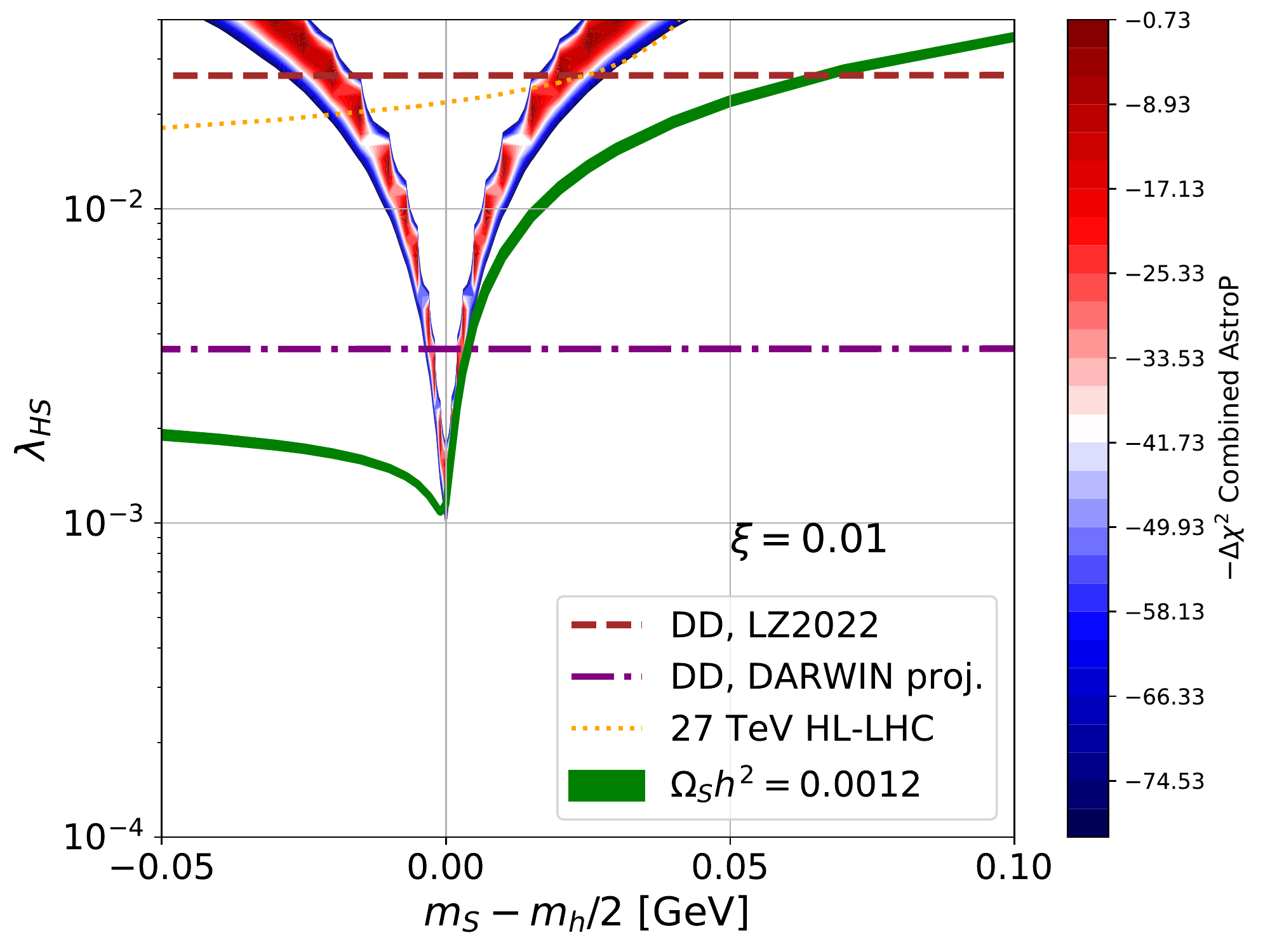}
\caption{The same as in Fig.~\ref{fig:combinedallMED} but for various values of the $\xi$ parameter, as labelled in the different frames.}    
\label{fig:Ssub}
\end{figure*}

\begin{figure*}[t]
\vspace{3.5ex}
\includegraphics[width=0.49\linewidth]{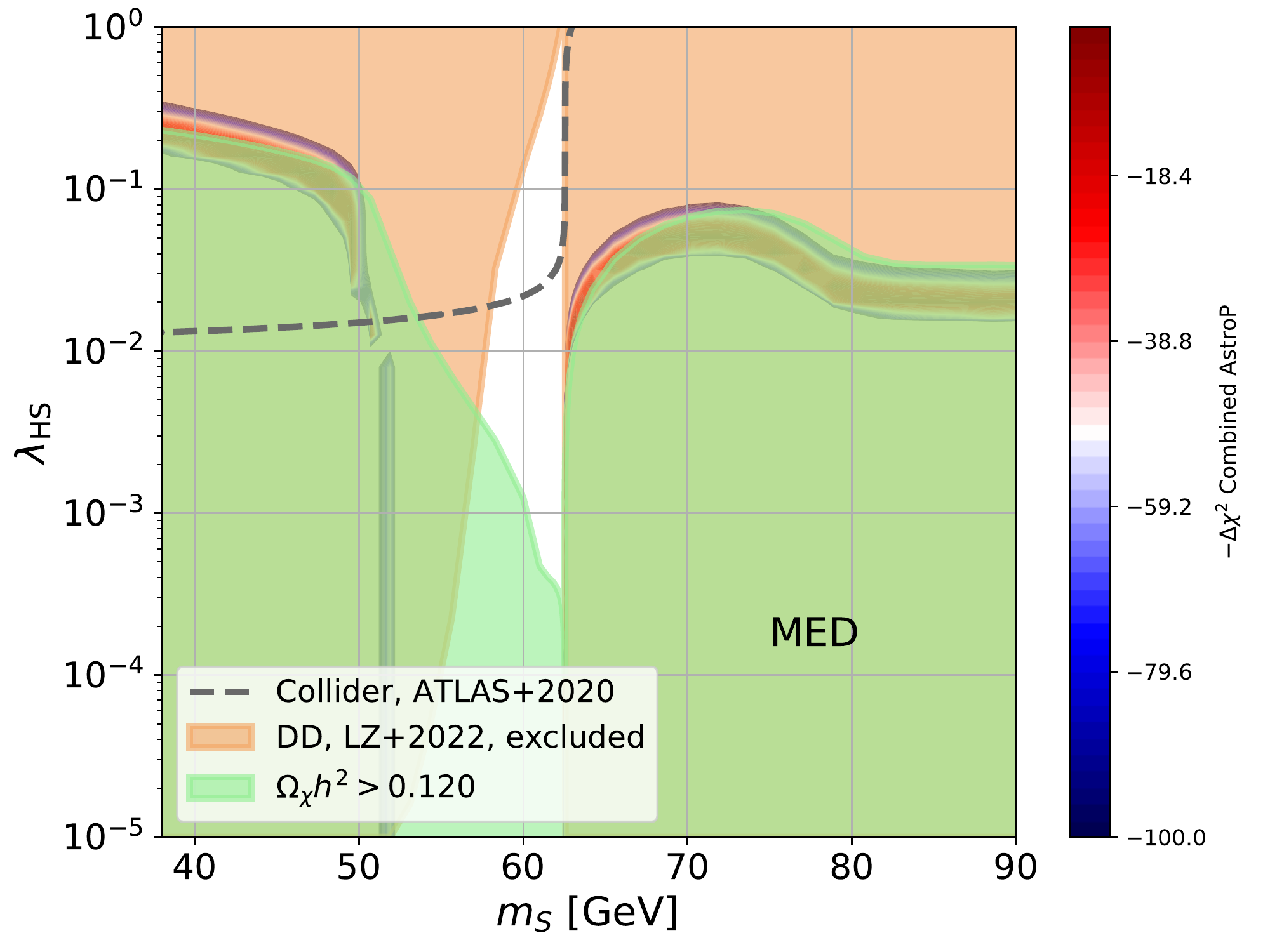}
\includegraphics[width=0.49\linewidth]{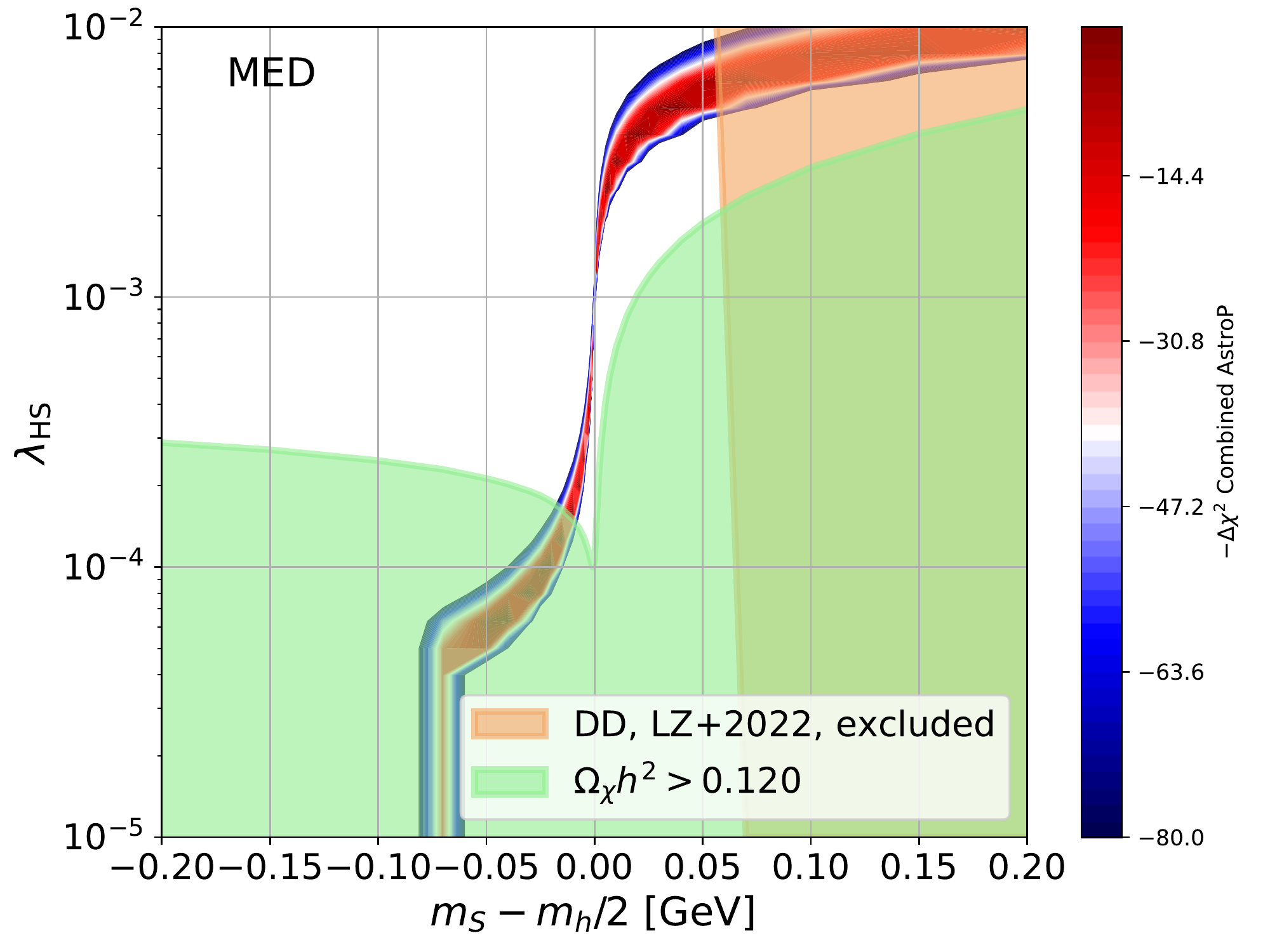}
\caption{Same as in Fig.~\ref{fig:combinedallMED} but rescaling, in each point of the parameter space, all predicted astrophysical signals (for direct and indirect detection) according to the DM fraction $\xi$, as it is predicted from the relic density computation. The green shading denotes the region with $\Omega h^2>0.12$ ($\xi>1$). The orange shaded region shows the  region excluded by the direct detection constraints from LZ data. The right panel show a zoom in the region of the resonance, better exposing the configurations compatible with all constraints.}
\label{fig:combinedxiMED}
\end{figure*}

\subsection{The \texorpdfstring{$S$}{S} particle as subdominant dark matter component}
\label{sec:Snotall}

In this section, we discuss the possibility that the $S$ particle is not the dominant component of DM density.
We therefore allow paramenter-space configurations for which $\Omega_Sh^2$ accounts from 1\% up 100\% of the measured dark matter average density $\Omega_{\rm DM}h^2$ (i.e. $0.01 \leq \xi \leq 1$).
We verify whether the relic density regions are still compatible with astroparticle data best-fit regions and with direct detection and collider constraints.
In Fig.~\ref{fig:RD1}, we show the parameter space of the SHP model providing a fraction of the DM relic density.
In order to find the best-fit region for $\lambda_{HS}$ and $m_S$ that fits cosmic fluxes data, we consider the following.
If the $S$ particle constitutes a fraction $\xi$ of the DM relic abundance, according to Eq.~\eqref{eq:relic_density_fraction}, the DM Galactic density should be rescaled by $\xi$.
The geometrical factor $\bar{\mathcal{J}}$ is proportional to $\rho^2$. This implies that $\bar{\mathcal{J}}\propto \xi^2$.
Since $\Phi\propto \bar{\mathcal{J}}\langle\sigma \vMol \rangle \propto \xi^2 \langle\sigma \vMol \rangle$, to obtain the same flux, $\langle\sigma \vMol \rangle$ has to increase  for decreasing $\xi$ according to $\langle\sigma \vMol \rangle \propto \xi^{-2}$. Since $\langle \sigma \vMol \rangle$, in turn, scales with $\lambda_{HS}^2$, for $\lambda_{HS}<0.1$ (see Fig.~\ref{fig:annih}), we can just rescale the values for $\lambda_{HS}$ for the combined fit to the cosmic particle flux data shown in the previous subsection by $1/\xi$.

\begin{figure}[t]
\includegraphics[width=1.00\linewidth]{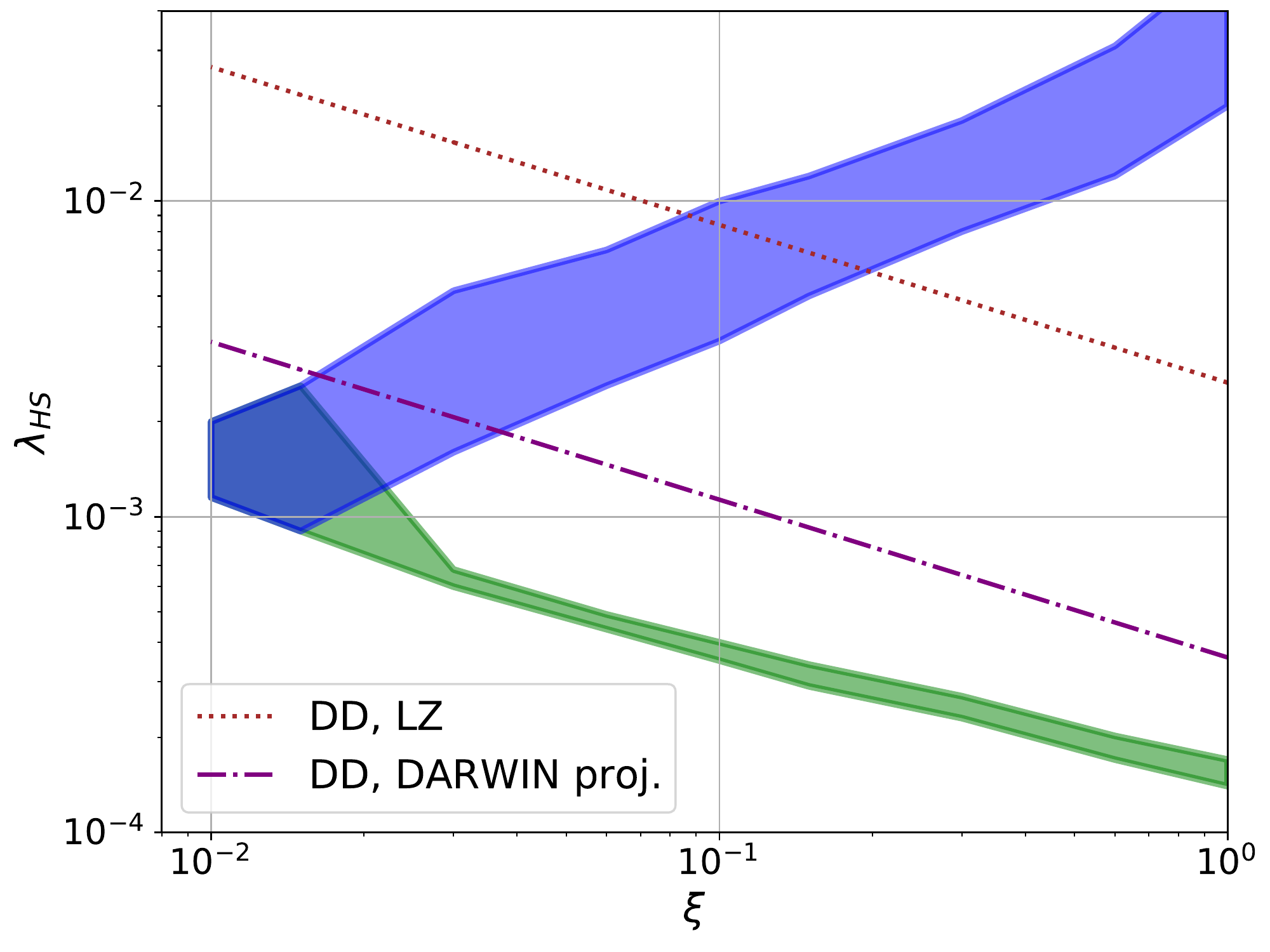}
\caption{Best-fit regions of the combined cosmic particle flux data compatible with thermal freeze-out (corresponding to the intersections of the green and color-shaded bands in the plots of Fig.~\ref{fig:Ssub}) in the $\xi$-$\lambda_{HS}$ plane. The green (blue) bands denote the regions where $M_S$ is slightly smaller (larger) than $m_h/2$. We also report the upper limits from direct detection covered by LZ and the future prospects for DARWIN.} 
\label{fig:Ssub_bis}
\end{figure}
In Fig.~\ref{fig:Ssub}, we show the results for the combined analysis for $\xi=0.3$, $0.1$, $0.03$, and $0.01$.
When $\xi = 0.3$ ($S$ makes up $30\%$ of total DM density) we see that the best-fit region to cosmic data matching the relic abundance corresponds to $m_S\approx m_h/2$ and $\lambda_{HS}=(2.3\!-\!2.7)\times 10^{-4}$, which is consistent with the direct detection upper limits. The second region with mass around $m_h/2+0.20$ GeV and $\lambda_{HS}=(0.8\!-\!1.8)\times 10^{-2}$ is instead excluded by LZ constraints.
For $\xi$ smaller than about $20\%$, also the higher mass region of compatibility between cosmic data and relic abundance becomes consistent with direct detection upper limits. This is visible in Fig.~\ref{fig:Ssub} for the case with $\xi=0.1$ where there is still a region of compatibility at about $m_S\approx m_h/2$ and $\lambda_{HS}\approx 3\times 10^{-4}$ and the second one which is at $m_S\approx m_h/2+0.05$ GeV has $\lambda_{HS}\approx (0.5\!-\!1.2)\times 10^{-2}$.
The smaller the value of $\xi$, the closer the high mass compatibility region is to the right side of $m_S\approx m_h/2$ with at the same time smaller values of $\lambda_{HS}$. For $\xi<0.02$ the two regions merge. This can be seen in Fig.~\ref{fig:Ssub_bis} showing the two best-fit regions of the cosmic particle flux data for $\lambda_{HS}$ as a function of $\xi$ demanding compatibility with thermal freeze-out (corresponding to the intersections of the green and color-shaded bands in the plots of Fig.~\ref{fig:Ssub}).
%showing the two best-fit regions in the $\xi$-$\lambda_{HS}$ plane.
For values of $\xi$ smaller than $1\%$ there is no region of the parameter space for which the best-fit from cosmic data is compatible with the relic density calculation. 
Therefore, if $\xi\geq0.20$ the only region fitting the cosmic data and compatible with the relic density and direct detection constraints is for $m_S\approx m_h/2$ with values of $\lambda_{HS}$ of the order of $(1.5\!-\!4)\times 10^{-4}$.
Instead, if $\xi = 0.01-0.20$, also slightly larger $m_S$ values are allowed, of the order of $0.05$ GeV larger, while the coupling can be as large as $10^{-2}$.

To summarise our results, in Fig.~\ref{fig:combinedxiMED}, we have plotted the best-fit region and constraints while rescaling all astrophysical signals according to the value of $\xi$ obtained from the relic density computation for each point in the parameter space. In the green shaded region, $\Omega_S h^2 > 0.12$, i.e.~this region would over-close the Universe and it is hence excluded. Above the shaded region, $\xi<1$ with $\xi$ decreasing towards larger $\lambda_{HS}$. The orange shaded region is excluded by direct detection constraint from LZ\@. As in Fig.~\ref{fig:combinedallMED}, the grey dashed line denotes the ($\xi$-independent) constraints from the LHC with the region above the line being excluded. In the right panel of Fig.~\ref{fig:combinedxiMED}, we zoom into the resonant region revealing the narrow parameter space preferred by the  observations.

The projected sensitivity of DARWIN is shown in Figs.~\ref{fig:Ssub} and~\ref{fig:Ssub_bis} as the purple dot-dashed curves. While the region with $m_S\lesssim m_h/2$ and smaller coupling remains out-of-reach for direct detection experiments in the foreseeable future,    the region with $m_S\gtrsim m_h/2$ (blue shaded area) can be probed with the sensitivity of DARWIN\@.

The results presented in this section are consistent with the ones reported in Ref.~\cite{Cuoco:2016jqt}.

\subsection{Results without the GCE and upper limits from cosmic antiprotons}
\label{sec:SHP_UL}

An alternative interpretation of the GCE states that it originates from the cumulative flux coming from a population of millisecond pulsars located in the Galactic bulge \cite{Bartels:2015aea,Lee:2015fea,Bartels:2017vsx,Macias:2016nev}.
If this is the case, tight constraints on a possible DM contribution in the Galactic center can be found (see, e.g.,~\cite{Abazajian:2020tww}).
Additionally, the analysis of a possible DM contribution to the cosmic ray flux of antiprotons have not found any signal, and upper limits on the annihilation cross section has been imposed in Ref.~\cite{Balan:2023lwg} with the model {\tt DIFF.BREAK}.

In this section, we report the results of our analysis assuming that the DM is not responsible for the GCE.
We take into account the results obtained by Ref.~\cite{Balan:2023lwg} with the model {\tt DIFF.BREAK}.
We can not use the results of Ref.~\cite{Abazajian:2020tww} since they do not consider the SHP model.

\begin{figure}[t]
\includegraphics[width=1.0\linewidth]{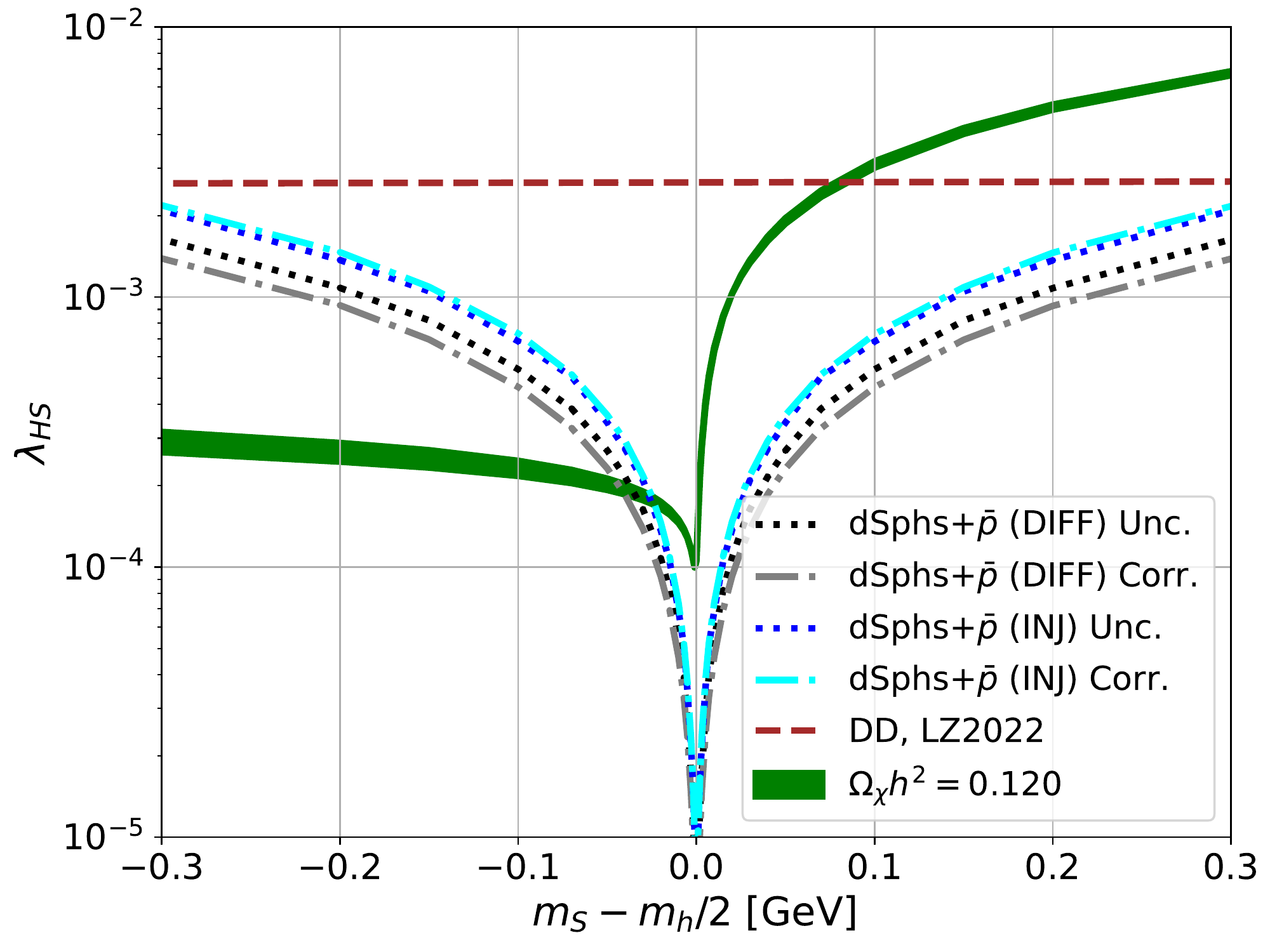}
\caption{Same as in Fig.~\ref{fig:combinedallMED} but with the upper limits derived from CR antiprotons flux data from AMS-02, with different assumptions for the propagation setup. In particular the model labed as `DIFF' (`INJ') stands for the model {\tt DIFF.BREAK} ({\tt INJ.BREAK}) in Ref.~\cite{Balan:2023lwg}. Moreover, the cases labeled as `Unc.' and `Corr.' refer to the case the uncertainties are taken to be uncorrelated and correlated, respectively.}
\label{fig:combinedUL}
\end{figure}

In Fig.~\ref{fig:combinedUL}, we show the results obtained with a combined analysis of dSphs, as presented in Sec.~\ref{sec:gamma}, and with antiprotons, following the results published in Ref.~\cite{Balan:2023lwg}.
We considered both {\tt DIFF.BREAK} and {\tt INJ.BREAK} models, with and without taking into account correlations between data points.
In particular, we display the $95\%$ CL upper limits for $\lambda_{HS}$ as a function of the DM mass obtained for the two propagation setups used in the antiproton analysis and the two cases of data correlations.
The upper limits rule out the region of the parameter space with $m_S\geq m_h/2$ because for those masses the values of $\lambda_{HS}$ which satisfy the relic density are above the upper limit curves.
Instead, for $m_S<m_h/2$ the allowed region contains masses below $m_h/2-0.05$ GeV and $\lambda_{HS}$ smaller than a few times $10^{-4}$.

\section{Conclusions}
\label{sec:conclusions}

In this paper, we performed a detailed and updated phenomenological study of the SHP model combining data from indirect and direct detection, the LHC, and the DM density measurement. We pinpointed the parameter space that is both allowed by the constraints imposed by this data as well as preferred as an explanation of the long-standing {\it Fermi}-LAT GCE and the AMS-02 antiproton excess. The former set of data alone leaves the resonant region $m_S\sim m_h/2$ as one of only two allowed windows of parameter space -- the other one being the high-mass region, $m_S\gtrsim 3$ TeV, which, however, requires large couplings $\lambda_{HS}\gtrsim1$ and will entirely be probed by future direct detection experiments, e.g.,~with DARWIN\@. In the resonant region, in contrast, very small couplings are sufficient to match the measured relic density containing a spot with maximal resonant enhancement that is out-of-reach for any direct detection and collider experiment in the foreseeable future. This leaves indirect detection as the only handle to conclusively probe this scenario.
 
Accordingly, we further narrow down the parameter space in the resonant region by testing its compatibility with the {\it Fermi}-LAT GCE and the AMS-02 antiproton excess independently pointing to a DM mass around $m_h/2$. We utilize recent results from Refs.~\cite{DiMauro:2021raz,Cholis:2021rpp} and \cite{Kahlhoefer:2021sha} based on 11 years of {\it Fermi}-LAT data and 7 years of AMS-02 data, respectively, thereby improving over previous results obtained within the model. Furthermore, special care has been taken in the precise computation of the relic density by considering effects of kinetic decoupling that are highly relevant close to the resonance. 

Two viable subregions fit both excesses while predicting the right relic density: one slightly below and one just above $m_h/2$ among which the latter is, however, excluded by direct detection due to the larger coupling required. 
When allowing $S$ to only constitute a fraction of the DM density -- while still being considered exclusively responsible for the observed indirect detection excesses -- more possibilities open up. In particular, due to the different scalings of indirect and direct detection rates with the DM fraction, the latter subregion becomes allowed for a fraction of 10\% or less. Both subregions provide good fits down to a DM fraction of 1\% below which the relic density and the indirect detection signals become incompatible. 

Providing couplings in between $10^{-2}$ and $10^{-4}$, both regions are extremely hard to probe with future experiments. We showed that a HL-LHC and even a 27 TeV HE-LHC could not test the model in the laboratory. However, with the sensitivity of DARWIN, the subregion with $m_S\gtrsim m_h/2$ and coupling $\lambda_{HS}\gtrsim 2\times 10^{-3}$ can be probed entirely.

A \python package enabling fast calculations of the dark matter spectra, relic density, as well as direct detection and collider constraints within the model is available on \href{https://github.com/dimauromattia/SingletScalar_DM}{\github}.

\begin{acknowledgments}
We thank Torsten Bringmann and Andrzej Hryczuk for their help in using the \drake code and Andreas Goudelis for providing information about \micromegas, Paolo Torrielli and Torbj\"orn Sj\"ostrand for the suggestions about the \pythia code and the authors of the Refs.~\cite{Kahlhoefer:2021sha,Balan:2023lwg} for providing the results of their papers.
N.F.~and M.D.M.~acknowledge support from the Research grant {\sl TAsP (Theoretical Astroparticle Physics)} funded by Istituto Nazionale di Fisica Nucleare (INFN). N.F. acknowledges support from: {\sl Departments of Excellence} grant awarded by the Italian Ministry of Education, University and Research (MIUR); Research grant {\sl The Dark Universe: A Synergic Multimessenger Approach}, Grant No. 2017X7X85K funded by the Italian Ministry of Education, University and Research (MUR). C.A. acknowledges support by the F.R.S.-FNRS under the “Excellence of Science” EOS be.h project no. 30820817. J.H.~acknowledges support by the Alexander von Humboldt foundation via the Feodor Lynen Research Fellowship for Experienced Researchers. 
\end{acknowledgments}

\bibliography{paper}

\clearpage
\newpage

\appendix

\section{Dark Matter energy spectra calculated with \maddm}
\label{app:spectrum}

In this section we provide the technical details for the calculation of the DM energy spectra with \maddm.
In particular, we refer to the computation of the quantities $\odv{N_p}/{E}$ with $p$ being $\gamma$ rays, positrons, antiprotons or neutrinos, that enter in the calculation of the flux of particles produced from DM annihilation (see Sec.~\ref{sec:ID}). 

As a first step, given a value of the DM mass $m_S$ and coupling $\lambda_{HS}$, \maddm calculates the cross section for all the annihilation channels.
Then, it obtains the relative contribution of each $i$-th channel as $\langle \sigma v \rangle_i/\langle \sigma v \rangle$, and it calculates the final spectrum of a certain particle $p$ as:
\begin{equation}
    \odv{N_p}{E} = \sum_i \Bigl(\odv{N_p}{E}\Bigr)_i \cdot \frac{\langle \sigma v \rangle_i}{\langle \sigma v \rangle_\textup{tot}} ,
\end{equation}
summing over the different annihilation channels labeled by the index $i$.
During the calculation of the annihilation cross sections, \maddm calls the {\tt MadEvent} generator to simulate events of the hard process, including the kinematics of the produced particles, eventually saving all the information in a Les Houches Event (LHE) file~\cite{Alwall:2006yp}.
Event numbers are generated proportional to the relative contributions of the channel $i$, $\langle \sigma v \rangle_i/\langle \sigma v \rangle_\textup{tot}$.

In the next step, \maddm internally calls the \pythia Monte Carlo generator, which takes as input the  LHE file. 
In particular, \pythia reads the LHE files and performs a Monte-Carlo simulation of the hadronization and showering of the DM annihilation final states, starting from the kinematics defined in the LHE file itself, and producing final energy spectra of stable particles, such as photons, neutrinos, $e^\pm$ and $\bar{p}$.
At this stage, loop-induced processes are handled differently: annihilation cross section calculation and related event generation, as well as the spectra simulation through \pythia, are performed independently and separately from tree-level processes.
Eventually, the spectra obtained from both the tree-level and loop-induced processes are combined according to their respective branching ratios.

The energy spectra we generate take into account FSR, which is included by \pythia using the command {\tt PartonLevel:FSR=on}, while we turn off initial-state-radiation ({\tt PartonLevel:ISR=off}), which is not present for DM annihilating particles, and multi-particle interactions ({\tt PartonLevel:MPI=off}).
We turn on the emission of weak gauge bosons off fermions which is part of FSR by using {\tt TimeShower:weakShower = on}.
Our procedure differs from the one used in the Ref.~\cite{Cirelli:2010xx} where the authors have generated DM spectra in a model independent framework. In particular, that approach implies the process is created through a fake resonance with energy equal to twice the DM mass, and making it decay into the channel of interest.
Some important effects are not taken into account if following this method.
The first is related to off-shell contributions from the massive gauge bosons. As we have seen in Fig.~\ref{fig:annih}, the contribution of off-shell $W^{\pm}$ and $Z$ could be very relevant between (60--90) GeV. 
The second effect is related to the polarization of $W^{\pm}$ and $Z$ bosons. By using the resonant mechanism as in Ref.~\cite{Cirelli:2010xx} the spin 1 gauge bosons are unpolarized, resulting in the final fermions produced after their decay to acquire the inaccurate kinematics. 
Instead, in our calculation, we include the polarization of the gauge bosons because, thanks to \maddm, we generate events up to the fermions produced from the decay of the $W^{\pm}$ and $Z$ particles. When \pythia reads the LHE files, the kinematics of the final fermions inherited the fact that the $W^{\pm}$ and $Z$ have a polarization.

\begin{figure*}[t]
\includegraphics[width=0.49\linewidth]{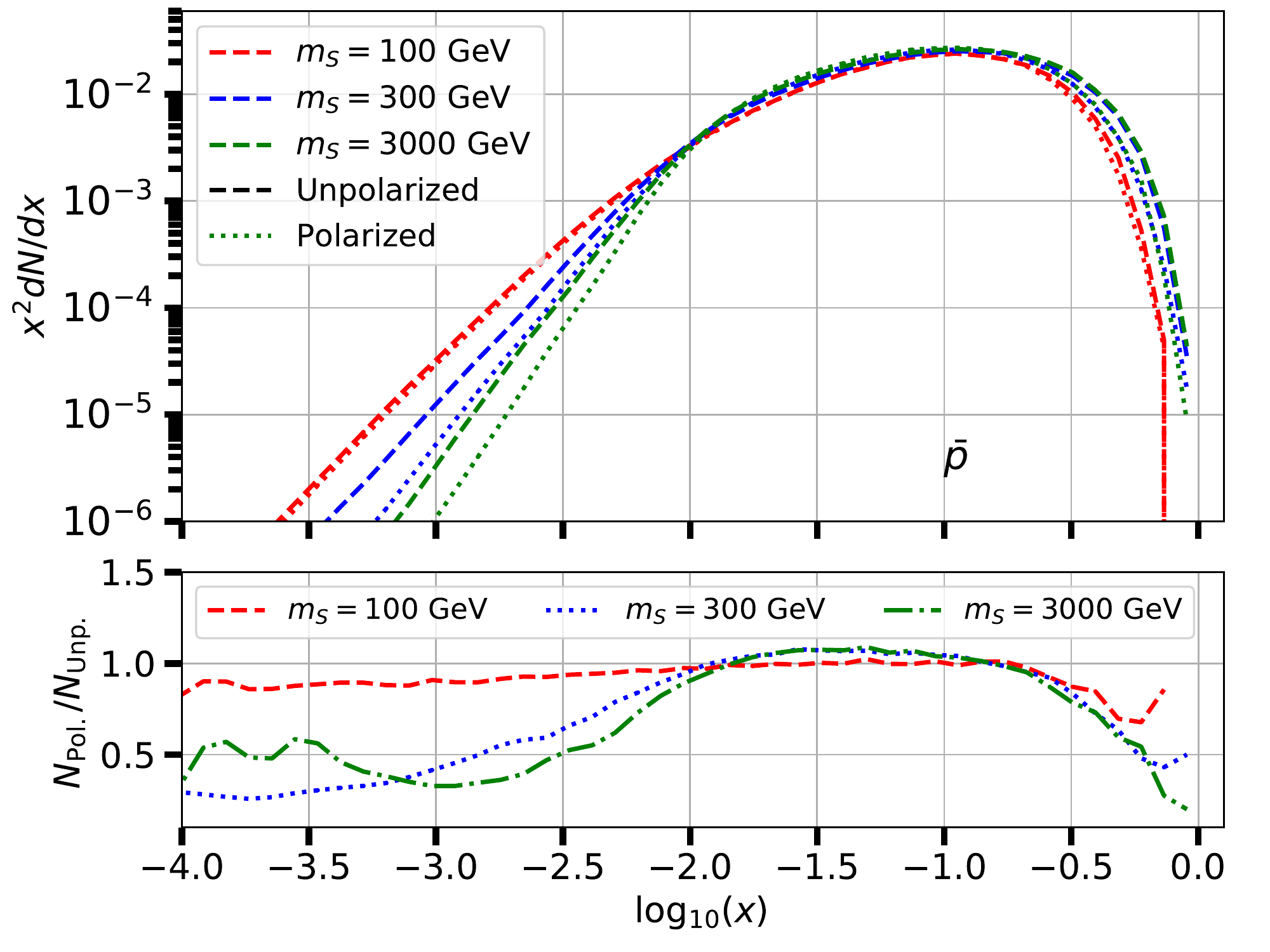}
\includegraphics[width=0.49\linewidth]{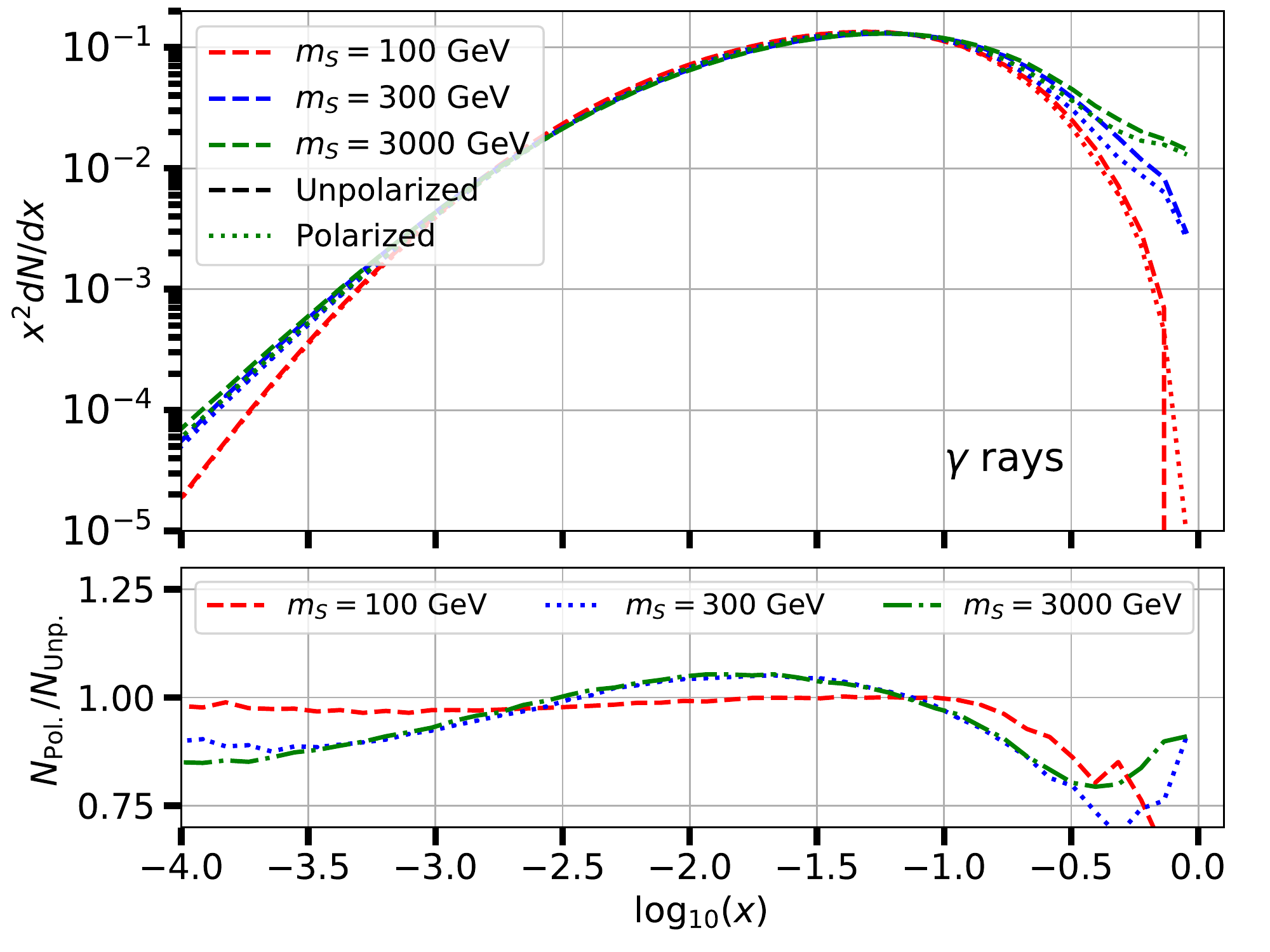}
\caption{Left panel: Spectra of $\bar{p}$ from DM annihilation in SHP model, reported as $x^2 \odv{N}{x}$, where $x=E/m_S$ and $E$ is the kinetic energy of the particles. Right panel: Same as left for the $\gamma$ rays. In the two panels we show three different values of $m_S$, for both unpolarized and polarized $W^{\pm}$ gauge bosons. In the bottom part of both plots we show the ratio between the polarized and unpolarized energy spectra.} 
\label{fig:DMspectraPol}
\end{figure*}

We show in Fig.~\ref{fig:DMspectraPol} the effect of the polarization of the $W^{\pm}$ and $Z$ bosons for the production of antiprotons and $\gamma$ rays.
In particular we consider three DM masses: 100, 300 GeV and 3 TeV.
The effect is more relevant for higher DM masses in the $\bar{p}$ channel, with respect to $\gamma$ rays.
In fact, for photons and $m_S=100$ GeV the effect accounts for a few $\%$ correction up to energies very close to the DM mass where the unpolarized spectra are larger by a factor between $(20\!-\!25)\%$.
Increasing $m_S$, the effect on the $\gamma$-ray spectra starts to be visible also at lower energies where the differences can be of the order of $(10\!-\!20)\%$.
Instead, for the production of $\bar{p}$, the effect could be larger and reaches even $(40\!-\!50)\%$ at small energies and large values of $m_S$.
In any case, the effect is small at the peak of the spectra, in units of $x^2 (\odv{N}/{x})$, where $x=E/m_S$, and increases in the low and high-energy tails.

\end{document}